\documentclass[sigconf,nonacm]{acmart}

\usepackage{balance}
\usepackage{verbatim}
\usepackage{amsmath,amssymb,amsfonts}

\usepackage{graphicx}
\usepackage{tabularx}
\usepackage{tikz}
\usetikzlibrary{positioning,shapes.geometric,arrows.meta}
\usepackage{subcaption}
\usepackage[export]{adjustbox}
\usepackage[utf8]{inputenc}
\usepackage{float}
\usepackage{svg}
\usepackage{enumitem}
\usepackage{booktabs}
\usepackage{threeparttable}
\usepackage{rotating}

\title{Fast and the Furious: Hot Starts in Pursuit-Evasion Games}  

\author{Gabriel Smithline}
\affiliation{%
  \institution{University of Michigan}
  \city{Ann Arbor}\state{MI}\country{USA}}
\email{gsmithl@umich.edu}

\author{Scott Nivison}
\affiliation{%
  \institution{Air Force Research Laboratory}
  \city{Shalimar}\state{FL}\country{USA}}
\email{scott.nivison.1@us.af.mil}

\begin{document}
\maketitle


\section{Introduction}
Pursuit-evasion games model interactions between a set of pursuers and evaders in differential games. These games are critically important in fields such as control, aerospace, and robotics, and form the foundation of adversarial optimal control \cite{isaacs1999differential, weintraub2020introduction}. Since their inception in the 1950s and 1960s, pursuit-evasion games have seen significant advancements, with one of the most foundational contributions being the Hamilton-Jacobi-Isaacs (HJI) equation \cite{isaacs1999differential} introduced by Rufus Isaacs. The HJI equation, a generalization of the Hamilton-Jacobi-Bellman (HJB) equation \cite{bardi1997optimal}, extends to zero-sum games where one player aims to minimize a cost function while another aims to maximize it. In these games, the optimal strategies for pursuers and evaders must satisfy the HJI. The equation, a partial differential equation (PDE), captures the game's dynamics, including player positions, velocities, and optimal control actions. The solution to the HJI equation, known as the value function, represents the minimum cost for pursuers to capture evaders, considering the evaders' optimal responses.

\begin{figure*}[t]
    \centering
    \setlength{\fboxrule}{1pt}  
    \begin{subfigure}[t]{0.45\linewidth}  
        \centering
        \fbox{\makebox[5cm][c]{\includegraphics[height=4cm]{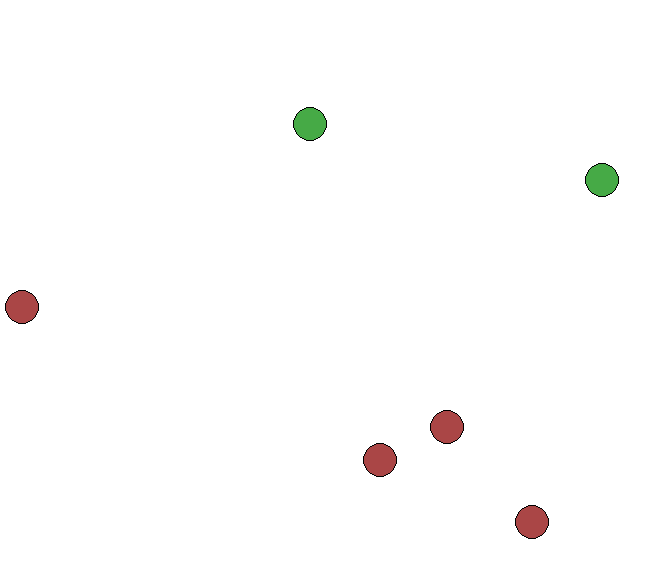}}}
        \caption{4 Pursuers vs. 2 Evaders Random Start}
        \Description{Image of random start.}
    \end{subfigure} \hfill
    \begin{subfigure}[t]{0.45\linewidth}
        \centering
        \fbox{\makebox[5cm][c]{\includegraphics[height=4cm]{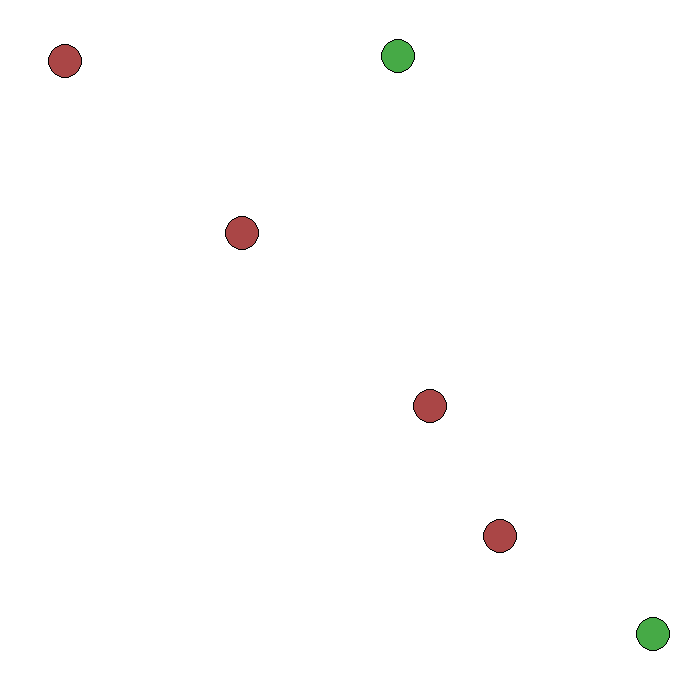}}}
        \caption{4 Pursuers vs. 2 Evaders Hot Start}
        \Description{Image of hot start.}
    \end{subfigure}
    \caption{Images of a hot start and a random start in 4 Pursuer vs. 2 Evader Game (pursuers are in red and evaders are in green).}
    \label{fig:4_v_2_example}
\end{figure*}

Despite these advancements, many open-ended questions remain. A significant challenge is strategically placing pursuers without knowing the number or locations of evaders. Developing methods to address this issue could enhance algorithm effectiveness and efficiency, thereby reducing collective pursuer costs.

In this paper, a novel approach is proposed that leverages artificial intelligence and machine learning techniques to offline generate initial configurations for pursuers before evaders appear, a strategy termed \textit{hot starts}. Inspired by recent advances in Graph Neural Networks (GNNs) \cite{scarselli2008graph}, particularly the methodology described in Google DeepMind's TacticAI developed for Liverpool FC \cite{wang2024tacticai}, we aim to apply similar methods to pursuit-evasion games. Two pure pursuit game scenarios are considered: One Evader vs. Many Pursuers and Many Evaders vs. Many Pursuers.

Our method demonstrates enhancements by not making direct assumptions about evader positions, quantities, algorithms, optimization issues, and state-space simplifications with these hot starts. To overcome the lack of domain-specific expert feedback on configurations, games are simulated using the Multi-Agent Particle Environment (MPE) \cite{lowe2017multi, mordatch2017emergence}, a flexible simulation framework designed for multi-agent interactions (see Figure~\ref{fig:pipeline_diagram}). Through these simulations, empirical data is generated, which serves two purposes: first, it provides the data needed to train the Graph Convolutional Network (GCN) \cite{kipf2016semi} to extrapolate patterns; second, it allows us to construct a Graph Feature Space (GFS) using three features: \( U_{\text{capture}} \) (capture potential), \( U_{\text{distance}} \) (distance), and \( U_{\text{heading}} \) (heading angle). Survival analysis and log-rank tests are performed to empirically evaluate whether hot starts have a statistically significant impact. 
Key contributions are outlined as follows:
\begin{itemize}
    \item Encode pursuer configurations in the GFS, allowing us to solve a multi-objective optimization problem offline and construct a Pareto frontier of pursuer configurations.
    \item Train the GCN using a loss function designed to learn from the Pareto-valued configurations. Track hypervolume and generational distance metrics to measure how the GCN learns the Pareto front and how well it generalizes by measuring the volume of the frontier it covers.
    \item Inspired by TacticAI, generate numerous configurations with a GCN and use statistical methods to evaluate the impact of these configurations on pursuer efficiency and evader capture.
\end{itemize}

\section{Related Work}
To model effective gameplay in pursuit-evasion games, algorithms often utilize Apollonius circles \cite{ramana2017pursuit, garcia2020multiple, isaacs1999differential}, and strategies based on Voronoi diagrams or Delaunay triangulation \cite{von2018pursuit, bakolas2010optimal} (for a more thorough overview on differential games, see Appendix Section A). Previous works have explored simplified, discretized environments where pursuers are explicitly assigned to evaders based on probabilistic estimation techniques, or where the explicit algorithms of the adversary are known by each party \cite{zhang2021pursuer, king2018informal, antoniades2003pursuit, giovannangeli2010pursuit}. Other studies highlight redundant allocation of pursuers \cite{von2018pursuit}, effectively increasing the cost of pursuit for the collective of pursuers.

Recent works have extended the application of graph-based learning techniques to multi-agent systems in discrete environments. For instance, Li et al. \cite{li2024grasper, li2023solving} structured agent interactions within a graph to solve pursuit-evasion games, with a focus on real-time coordination and task assignment. While effective for discrete settings, these methods do not address the dynamic and continuous aspects of real-world environments, nor do they tackle the same problem focused on in this work. This work diverges by introducing a novel process for generating "hot starts" without prior knowledge of evader positions. By leveraging differential game theory and operating in continuous space, time-dependent algorithms are accounted for, for both pursuers and evaders, which contrasts with the real-time updates used in discrete environments. This multi-objective optimization approach allows us to generate Pareto-optimal configurations that improve pursuer efficiency from the outset. The combination of continuous space, differential dynamics, and the distinct focus on initial agent allocation sets this work apart in the domain of pursuit-evasion games, making it applicable to a wider range of real-world scenarios.

The concept of initial agent allocation in large-scale systems has also been explored in particle swarm optimization and genetic algorithms, particularly in population initialization, where the goal is to initialize the population advantageously \cite{kazimipour2014review}. Traditionally, evolutionary algorithms use pseudo-random number generators for population initialization; more recent work explores intelligent initialization through AI-based heuristics \cite{8616132}.

While GNN-based approaches like those used in Google DeepMind's TacticAI for sports strategy \cite{wang2024tacticai} have shown success in generating player configurations for specific tasks (e.g., corner kicks), applying similar methods to pursuit-evasion games poses distinct challenges. In sports, configurations benefit from domain-specific expertise and cl-ear objectives, such as the positioning of players based on real-time feedback from coaches. By contrast, pursuit-evasion games often lack this level of expert guidance, and there is the added complexity of not knowing evader positions or quantities in advance.

Previous works have not fully explored the challenge of strategically positioning pursuers without such prior knowledge, which is a fundamental aspect of real-world pursuit-evasion scenarios. This approach seeks to explore this gap by leveraging GNNs to generate initial configurations, termed "hot starts," without making direct assumptions about the number or location of evaders. This method has potential applications in a wide range of domains, including UAV search and rescue \cite{furukawa2006recursive}, wildlife protection \cite{xu2020stay, bondi2018spot}, human-computer interaction \cite{li2019differential}, traffic management \cite{9661291}, and general security \cite{annas2009intelligent, tambe2011security}, where such strategic initialization could significantly enhance operational efficiency.
\begin{figure}[H]
    \centering
\includegraphics[width=0.8\linewidth]{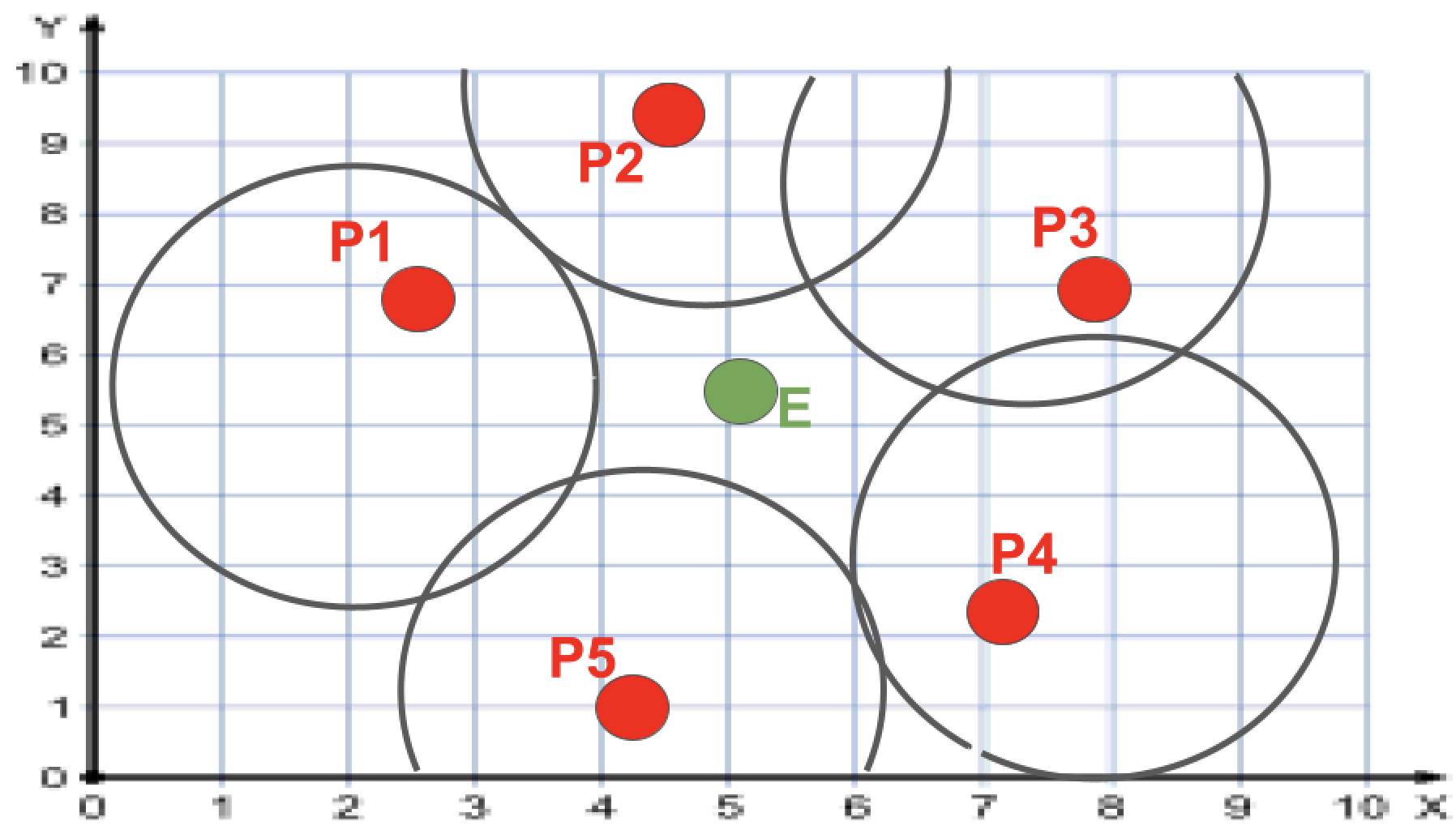}
    \caption{Image of 5 pursuers (in red) surrounding an evader (in green), illustrating cooperative containment.  Black circles represent capture radii.}
    \label{fig:capture_raddi_example}
    \Description{Diagram showing 5 pursuers (in red) surrounding an evader (in green). The links $\theta_{i, j}$ between the pursuers are strictly positive, illustrating cooperative containment. Black circles represent capture radii.}
\end{figure}
\section{Control Algorithms}

\subsection{1 Evader vs. Many Pursuers}
Control algorithms using heuristics inspired by strategies from \cite{garcia2020multiple, 9483097} were adapted for our scenarios, taking into account the varying speed ratios between evaders and pursuers. In the literature, different speed ratios lead to distinct geometric constructs: Apollonius circles or Cartesian ovals when $\gamma \neq 1$, and perpendicular bisectors or hyperbolas when $\gamma = 1$. In our work, for the 1 evader vs. many pursuers scenario, we assume $\gamma = \frac{v_E}{v_P} > 1$, where the evader is faster than the pursuers. Here, the evader identifies the weakest link between pursuers to escape, while pursuers aim to contain the evader strategically. (See Appendix B for detailed algorithms and system dynamics.)

\subsection{Many Evaders vs. Many Pursuers}
For many-vs-many scenarios ($\gamma = \frac{1}{2} < 1$), where pursuers are faster than evaders, we employ a hybrid approach combining geometry-based controls with PDE solutions. Specifically, we use a first-order upwind finite difference scheme with directional bias: backward differences are selected for negative gradients and forward differences for positive gradients. This scheme provides numerical stability while respecting the information flow direction in the HJI equation, consistent with methods described in \cite{Mitchell2005TimeDependentHJ, Osher2003LevelSM}. The Hamiltonian computation explicitly accounts for the minimizing pursuer and maximizing evader roles, ensuring appropriate upwind direction selection based on agent type.

Our hybrid approach uses a weighted parameter $\alpha = 0.5$ to combine geometric distance metrics with PDE value function information in the assignment cost matrix for the Hungarian algorithm. The geometric measures specifically include: (1) Euclidean distances between pursuers and evaders for the assignment cost matrix, (2) normalized direction vectors for determining movement trajectories, (3) relative bearing angles for optimal interception paths, and (4) proximity-based avoidance vectors that allow evaders to compute escape directions from nearby pursuers. Pursuers are allocated to evaders by minimizing this hybrid cost, and control actions are computed using both gradient information from the PDE and direct geometric vectors, combining them as:
\begin{equation}
\mathbf{u}_{\text{final}} = \alpha \cdot \mathbf{u}_{\text{geometric}} + (1-\alpha) \cdot \mathbf{u}_{\text{gradient}}
\end{equation}
where $\mathbf{u}_{\text{geometric}}$ is derived from the instantaneous geometric relationship between agents, and $\mathbf{u}_{\text{gradient}}$ comes from the spatial derivatives of the value function. Evaders employ escape strategies based on gradient ascent of the value function combined with geometric avoidance of nearby pursuers, computing avoidance vectors that are inversely proportional to the squared distance from approaching pursuers when they enter a predefined avoidance radius. 

Our numerical approach, while using discrete time steps, aims to approximate continuous-time game dynamics. The qualitative results align with theoretical predictions for pursuit-evasion scenarios with the given speed ratios \cite{isaacs1969differential, bacsar1998dynamic, weintraub2020introduction}, demonstrating the efficacy of our implementation despite the inherent limitations of numerical approximation.

\begin{figure*}[t]
    \centering
    \resizebox{0.75\textwidth}{!}{\includegraphics{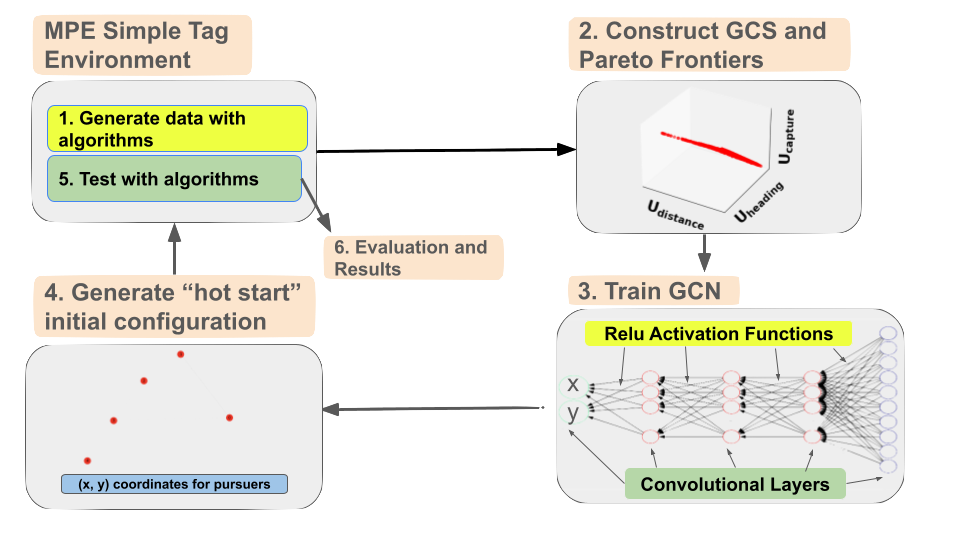}} 
    \caption{Pipeline diagram of the Pursuit-Evasion Games framework. Starting from the MPE Simple Tag Environment, the process follows these steps: 1. Generate data using control algorithms; 2. Construct the Graph Feature Space (GFS) and identify Pareto frontiers; 3. Train the Graph Convolutional Network (GCN); 4. Generate Hot Starts (each red dot represents x and y coordinates); 5. Apply these Hot Starts to the control algorithm; and 6. Perform statistical evaluation.}
    \label{fig:pipeline_diagram}
    \Description{A pipeline diagram representing the Pursuit-Evasion Games framework. The diagram outlines the following steps: starting from the MPE Simple Tag environment, data is generated using control algorithms. The next step involves constructing the Graph Feature Space (GFS) and identifying Pareto frontiers. Following that, the Graph Convolutional Network (GCN) is trained, and hot starts, represented by red dots showing x and y coordinates, are generated. These hot starts are then applied to the control algorithms, followed by statistical evaluation.}
\end{figure*}
\balance
\section{Graph Feature Space}

\subsection{Graph Features}

The Graph Feature Space (GFS) is an objective space where configurations of agents are mapped based on certain features, sampled during our data-generation phase. The three features we focus on to construct the Pareto frontier are: \( U_{\text{capture}} \) (capture potential), \( U_{\text{distance}} \) (distance), and \( U_{\text{heading}} \) (heading angle). These features are computed offline for each pursuer within the PettingZoo environment, as illustrated in Figure~\ref{fig:pipeline_diagram}. We delve into their definitions below.

\subsection{Pareto Frontier Definition}

The Pareto frontier is formally defined as the subset \( \Pi_p \subset \Pi \), where \( \Pi \) is the set of all possible configurations. Formally, \( \Pi_p \) is found such that:
\begin{equation}
\Pi_p = \{ G_i \in \Pi \mid \nexists G_j \in \Pi \text{ such that } \hat{c}_{G_j} \succ \hat{c}_{G_i} \},
\end{equation}
where \( \hat{c}_{G_i} \) is the vector containing our three features of the GFS, for configuration \( G_i \), and \( \hat{c}_{G_j} \succ \hat{c}_{G_i} \) indicates that the feature vector of \( G_j \) strictly dominates that of \( G_i \). More precisely, for \( G_j \) to strictly dominate \( G_i \):
\begin{align}
\hat{c}_{G_j} \succ \hat{c}_{G_i} &\iff \forall k \in \{1, \dots, n\}, \hat{c}_{G_j,k} \geq \hat{c}_{G_i,k} \nonumber\\
&\quad \wedge \exists l \in \{1, \dots, n\}, \hat{c}_{G_j,l} > \hat{c}_{G_i,l}
\end{align}
where \( n \) is the number of objectives. The set \( \Pi_p \) thus represents the Pareto-optimal configurations, where no configuration in \( \Pi \) strictly dominates any element of \( \Pi_p \) \cite{Zheng2024BridgingTG}.

\subsection{Feature Definitions}

Specifically, \( U_{\text{capture}} \) is expressed as:
\begin{equation}
U_{\text{capture}} = 2 \left(1 - \frac{1}{1 + e^{\frac{-d_i + r}{r}}}\right)
\end{equation}
where:
\begin{itemize}
    \item \( d_i \) is the distance between the pursuer and the \( i \)-th evader.
    \item \( r \) is the capture radius.
    \item \( n \) is the number of evaders.
\end{itemize}
This utility is designed to be 1 when the distance \( d_i = r \), and approaches 0 as the distance \( d_i \) increases. When \( d_i < r \), \( U_{\text{capture}} > 1 \), indicating a higher capture potential within the capture radius. The sigmoid component smoothly transitions the utility value, avoiding abrupt changes and adding stability to optimization and decision-making processes.

\( U_{\text{distance}} \) is expressed as:
\begin{equation}
U_{\text{distance}} = \min_{i=1, \ldots, n} \| \mathbf{p} - \mathbf{e}_i \|
\end{equation}
where:
\begin{itemize}
    \item \( \mathbf{p} \) is the position vector of the pursuer.
    \item \( \mathbf{e}_i \) is the position vector of the \( i \)-th evader.
    \item \( n \) is the number of evaders.
\end{itemize}
This utility represents the shortest distance between the pursuer and any of the evaders. It focuses on the closest evader, which is critical in pursuit-evasion games where the goal is to minimize the distance to capture the nearest evader.

Finally, \( U_{\text{heading}} \) is expressed as:
\begin{equation}
U_{\text{heading}} = \frac{1}{n} \sum_{i=1}^{n} \arccos \left( \frac{\mathbf{v}_p \cdot (\mathbf{e}_i - \mathbf{p})}{\| \mathbf{v}_p \| \| \mathbf{e}_i - \mathbf{p} \| + \epsilon} \right)
\end{equation}
where:
\begin{itemize}
    \item \( \mathbf{v}_p \) is the velocity vector of the pursuer.
    \item \( \mathbf{p} \) is the position vector of the pursuer.
    \item \( \mathbf{e}_i \) is the position vector of the \( i \)-th evader.
    \item \( n \) is the number of evaders.
    \item \( \epsilon = 10^{-9} \) is a small constant to prevent division by zero.
\end{itemize}
This utility measures the alignment between the pursuer's heading direction and the directions to each evader. Lower values of \( U_{\text{heading}} \) (i.e., smaller angles) indicate better alignment between a pursuer's heading and the direction to an evader when locations are known. In scenarios where evader locations are unknown, maximizing \( U_{\text{heading}} \) (achieved by averaging) encourages pursuers to consider a wider range of heading angles, increasing the likelihood that pursuers are heading in beneficial directions when evaders appear.

Illustrative examples demonstrating the functionality of $ U_{\text{heading}}$ and $U_{\text{capture}}$ are provided in Section D of the Appendix.  

\subsection{Normalization and Aggregation}

To ensure comparability among the different features and maintain the sum-to-one constraint, we normalize the aggregated features proportionally. Specifically, for each configuration \( G_i \), we aggregate the utilities by averaging over all pursuers:
\[
U_k(G_i) = \frac{1}{N_p} \sum_{j=1}^{N_p} U_k^{(j)}, \quad k \in \{\text{capture}, \text{distance}, \text{heading}\}
\]
where \( N_p \) is the number of pursuers in the configuration \( G_i \), and \( U_k^{(j)} \) is the feature for pursuer \( j \).

To enforce the sum-to-one constraint across all utilities, each aggregated utility \( U_k(G_i) \) is normalized by the total sum of all three utilities for the configuration:
\[
\tilde{U}_k(G_i) = \frac{U_k(G_i)}{\sum_{m=1}^{3} U_m(G_i)}
\]
This ensures that:
\[
\sum_{k=1}^{3} \tilde{U}_k(G_i) = 1
\]
By applying this proportional normalization, we maintain a balanced contribution of each utility feature in the multi-objective optimization process. This facilitates a fair and consistent comparison of pursuer configurations, ensuring that no single feature disproportionately influences the optimization outcomes.

\subsection{Construction of the Graph Feature Space (GFS)}
    
To construct the GFS and identify Pareto-optimal pursuer configurations, we formulate a constrained multi-objective optimization problem. We utilize the NSGA-II genetic algorithm \cite{deb2002fast}, with a population size of 500 and 200 generations, to solve this problem. We start by generating a diverse set of pursuer configurations. Each configuration \( G_i \) consists of multiple pursuers positioned in the environment. These configurations are generated through simulations or sampling methods and are stored in the set \( \Pi \).

For each configuration \( G_i \in \Pi \), we compute the three normalized and aggregated features:
\[
\tilde{U}_k(G_i) = \frac{U_k(G_i)}{\sum_{m=1}^{3} U_m(G_i)}, \quad k \in \{\text{capture}, \text{distance}, \text{heading}\}
\]
    
Our goal is to identify configurations that offer the best trade-offs among the three features. We formulate the multi-objective optimization problem as:
\begin{align}
\underset{G_i \in \Pi}{\text{minimize}} & \quad \mathbf{f}(G_i) = \left[ -\tilde{U}_{\text{capture}}(G_i), \tilde{U}_{\text{distance}}(G_i), -\tilde{U}_{\text{heading}}(G_i) \right]
\end{align}
    
The objectives are:
\begin{itemize}
    \item \textbf{Maximize \( \tilde{U}_{\text{capture}}(G_i) \)}: Increase the potential effectiveness in capturing evaders.
    \item \textbf{Minimize \( \tilde{U}_{\text{distance}}(G_i) \)}: Reduce the average minimum distance to evaders.
    \item \textbf{Maximize \( \tilde{U}_{\text{heading}}(G_i) \)}: Encourage pursuers to have diverse heading directions.
\end{itemize}
    
Since standard optimization algorithms perform minimization, we negate the objectives we wish to maximize. The proportional normalization ensures that each feature is appropriately scaled, facilitating a balanced optimization process.
    
From the Pareto-optimal configurations determined by our optimization problem, we construct graphs by adding edges between pursuer nodes called PP Edges.

\subsubsection{Pursuer-Pursuer (PP) Edges}

A PP edge exists between two pursuers if the distance between them is less than or equal to the average distance between all pairs of pursuers in the configuration. The weight of a PP edge is defined as:
\begin{equation}
w_{\text{PP}} = \frac{1}{d_{pp}}
\end{equation}
where \( d_{pp} \) is the distance between the two pursuers.

Each node in the graph consists of the following features: \( \tilde{U}_{\text{capture}} \), \( \tilde{U}_{\text{distance}} \), \( \tilde{U}_{\text{heading}} \), capture radius, \( x \) \& \( y \) velocities, and \( x \) \& \( y \) positions.

We employ the NSGA-II algorithm to solve this optimization problem. The algorithm operates as follows:
\begin{enumerate}
    \item \textbf{Decision Variables}: The configurations \( G_i \) in \( \Pi \).
    \item \textbf{Evaluation}: For each configuration \( G_i \), compute the normalized features and evaluate the objective vector \( \mathbf{f}(G_i) \).
    \item \textbf{Constraint Handling}: Enforce the constraint that the sum of the features equals 1.
    \item \textbf{Selection}: Identify non-dominated configurations to form the Pareto front.
\end{enumerate}

The result of the optimization is a set of Pareto-optimal configurations \( \Pi_p \subset \Pi \). These configurations represent the best trade-offs among the objectives, where no other configuration in \( \Pi \) is strictly better in all objectives.

\section{Graph Convolutional Network}

To generate our hot starts, we use a Graph Convolutional Network (GCN) trained on the graphs from the GFS that fall on the Pareto frontier. GCNs are advantageous for operating on graph-structured data, making them suitable for tasks like node classification, link prediction, and graph generation. The GCN follows a structure where it processes the features of pursuer nodes and their connections to output optimized configurations.

\subsection{Pareto Loss}

During training, we monitor several performance measures such as Hypervolume, Generational Distance (GD), Generational Distance Plus (GD+), Inverted Generational Distance (IGD), and Inverted Generational Distance Plus (IGD+) \cite{pymoo,zitzler2007hypervolume}. Our primary loss function, called Pareto loss, is defined to measure the distance between the utility feature vector of the generated pursuer configuration and the utility feature vectors of the configurations on the Pareto front:
\begin{equation}
\mathcal{L}_{\text{Pareto}} = \min_{j \in \{1, \dots, m\}} \left\| \mathbf{y} - \mathbf{p}_j \right\|
\end{equation}
where:
\begin{itemize}
    \item \( \mathbf{y} \) is the aggregated utility feature vector for the generated pursuer configuration, computed by averaging the utilities of all pursuers in the configuration:
    \[
    \mathbf{y} = \frac{1}{n} \sum_{i=1}^{n} \left( \tilde{U}_{\text{distance}_i}, \tilde{U}_{\text{capture}_i}, \tilde{U}_{\text{heading}_i} \right)
    \]
    where \( n \) is the number of pursuers.
    \item \( \mathbf{p}_j \) is the utility feature vector of the \( j \)-th configuration on the Pareto front.
    \item \( m \) is the number of configurations on the Pareto front.
\end{itemize}
This loss function encourages the network to generate configurations whose aggregated utility vectors are close to those of the Pareto-optimal configurations.

\subsection{Network Layers and Training}
Our GCN processes the features of all pursuers in a configuration simultaneously, considering the interactions between pursuers through the graph structure. The input features for each pursuer node include:
\begin{itemize}
    \item \textbf{Pursuer Position (\( x, y \))}: The initial positions of the pursuers.
    \item \textbf{Pursuer Velocity (\( v_x, v_y \))}: The initial velocities or headings of the pursuers.
    \item \textbf{Additional Features}: Such as initial utility estimates and capture radius or apollonious circle.
\end{itemize}
The edges in the graph represent connections between pursuers (PP edges), allowing the GCN to aggregate information from neighboring nodes and consider their relative positions and features. 

The network outputs adjusted positions and velocities for all pursuers simultaneously, ensuring that the generated configuration is coherent and optimized at the group level. By processing the entire configuration as a graph, the GCN accounts for interactions between pursuers, preventing issues such as overlapping positions and promoting effective spatial distribution.

Our network architecture and training parameters are detailed in Table~\ref{tab:combined-table}. Training took place on 4,700 graphs from 12 game types (combinations of 1 to 5 evaders and 2 to 5 pursuers), effectively stacking the Pareto fronts of each game type to form the GFS. GD, GD+, IGD, and IGD+ were measured during training to assess how the generated graphs relate to our Pareto values. The hypervolume of the generated graphs was approximately 0.634 (see Appendix D). Hypervolume was not included in the loss function due to computational expense.

Although training ran for 150 epochs, the network converged earlier, as loss values decreased significantly over time. Each utility feature ranges from 0 to 1; thus, small changes in the GFS led to larger changes in actual results. Despite early convergence, extended training was necessary to fine-tune the network, as it was found that small changes in the GFS led to larger changes in actual results.

\begin{table}[H]
\vspace{-2mm}
\centering
\caption{Network Architecture and Loss Values, Training Parameters}
\label{tab:combined-table}
\begin{minipage}[t]{0.44\linewidth}
\centering
\textbf{(a) Network Architecture and Loss Values}

\begin{tabular}{@{}p{1.2cm}p{1cm}p{0.6cm}@{}}
\toprule
\textbf{Layer} & \textbf{Nodes} & \textbf{Act.} \\
\midrule
Input          & 10             & ReLU          \\
Hidden 1       & 64             & ReLU          \\
Hidden 2       & 32             & ReLU          \\
Hidden 3       & 64             & ReLU          \\
Output         & 2              & Linear        \\
\bottomrule
\end{tabular}
\vspace{3mm}
\begin{tabular}{@{}p{1cm}p{1cm}@{}}
\toprule
\textbf{Epoch} & \textbf{Loss} \\
\midrule
1              & 0.145        \\
75             & 0.0247        \\
150            & 0.0234        \\
\bottomrule
\end{tabular}
\end{minipage}%
\hspace{2mm}
\begin{minipage}[t]{0.44\linewidth}
\vspace{1mm}
\centering
\textbf{(b) Training Parameters}
\begin{tabular}{@{}p{2.6cm}p{1.2cm}@{}}
\toprule
\textbf{Parameter} & \textbf{Value} \\
\midrule
Batch Size         & 1024           \\
Learning Rate      & $1\times10^{-4}$ \\
Optimizer          & Adam           \\
Weight Decay       & $1\times10^{-2}$ \\
Epochs             & 150            \\
Hardware           & L4 GPU         \\
\bottomrule
\end{tabular}
\end{minipage}
\end{table}

\section{Experiments and Discussion}

After training, our network generated 1000 configurations for each game type. These configurations were generated offline without prior knowledge of evader locations or numbers, then stored in a CSV, and supplied as initial pursuer locations as opposed to randomly placing them on a plane following a uniform distribution. To query our network, we passed in the amount of pursuers we would like it to generate configurations for (this could be thought of as the game type), as well as random input for the other features, meant to signify noise; this is meant to capture that fact that we have no observability about their velocities or locations. The network relies solely on pursuer-related inputs, producing positions and headings that are likely to perform well against unknown evader placements. This approach is distinct and fundamentally unique, as it enables the generation of strategically advantageous pursuer configurations without requiring real-time evader information.

We randomly sampled 10 batches of 100 configurations for each game type and tested them, comparing results with configurations generated using random placements. The results demonstrate that our network effectively learns generalizable configurations that improve performance in pursuit-evasion games, even without specific evader information.


\begin{table}[tb]  
\centering
\caption{Survival Probabilities for 4 Pursuers vs.\ 2 Evaders (GCN vs.\ Random).}
\label{tab:survival_probabilities_two_col}
\resizebox{\columnwidth}{!}{%
\begin{tabular}{@{}r r r r r r@{}}
\toprule
\textbf{Time} & \textbf{GCN} & \textbf{Rand} & 
\textbf{Time} & \textbf{GCN} & \textbf{Rand} \\
\midrule
0  & 1.0000 & 1.0000 & 21 & 0.8722 & 0.9133 \\
1  & 0.9991 & 0.9984 & 22 & 0.8619 & 0.9061 \\
2  & 0.9981 & 0.9958 & 23 & 0.8511 & 0.8984 \\
3  & 0.9967 & 0.9931 & 24 & 0.8398 & 0.8903 \\
4  & 0.9943 & 0.9902 & 25 & 0.8279 & 0.8817 \\
5  & 0.9909 & 0.9873 & 26 & 0.8154 & 0.8725 \\
6  & 0.9865 & 0.9842 & 27 & 0.8022 & 0.8627 \\
7  & 0.9812 & 0.9811 & 28 & 0.7881 & 0.8522 \\
8  & 0.9753 & 0.9777 & 29 & 0.7732 & 0.8409 \\
9  & 0.9691 & 0.9742 & 30 & 0.7572 & 0.8285 \\
10 & 0.9626 & 0.9705 & 31 & 0.7398 & 0.8149 \\
11 & 0.9559 & 0.9667 & 32 & 0.7210 & 0.7998 \\
12 & 0.9488 & 0.9626 & 33 & 0.7002 & 0.7829 \\
13 & 0.9415 & 0.9583 & 34 & 0.6772 & 0.7637 \\
14 & 0.9339 & 0.9537 & 35 & 0.6511 & 0.7418 \\
15 & 0.9260 & 0.9489 & 36 & 0.6208 & 0.7160 \\
16 & 0.9179 & 0.9438 & 37 & 0.5843 & 0.6848 \\
17 & 0.9094 & 0.9384 & 38 & 0.5385 & 0.6448 \\
18 & 0.9007 & 0.9327 & 39 & 0.4753 & 0.5882 \\
19 & 0.8915 & 0.9266 & 40 & 0.3635 & 0.4846 \\
20 & 0.8821 & 0.9201 &  -- &   --   &   --   \\
\bottomrule
\end{tabular}%
} 
\vspace{2pt} 
\footnotesize
\textit{Note}: Times 0--20 are in the left half, 21--40 in the right half. 
Survival probabilities estimated via Kaplan-Meier at each discrete time step.
\end{table}

\subsection{Methodology}
Hot starts are evaluated using a Monte Carlo analysis, aggregating results across numerous simulations. One can think of random starts as a control group and hot starts as the treatment group. Then the  Kaplan-Meier estimator, a nonparametric metric from biostatistics \cite{rich2010practical, goel2010understanding, kaplan1958nonparametric}, was used to estimate evader survival probabilities over time. Log-rank tests were performed to validate the statistical significance of our results.  

\begin{figure*}[t]
    \centering
    \begin{subfigure}[t]{0.45\linewidth}  
        \centering
        \includegraphics[height=4cm]{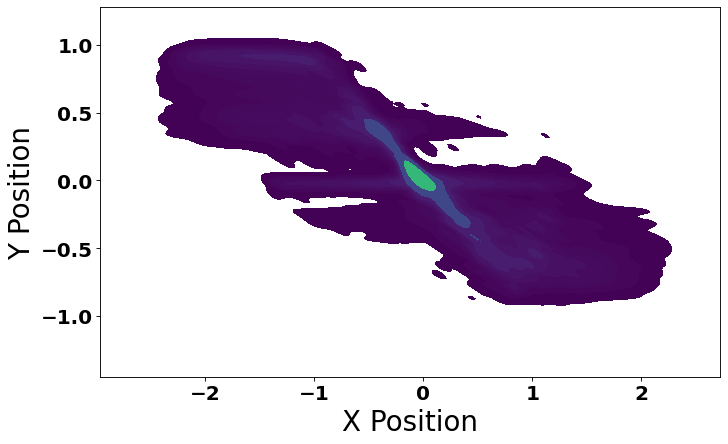}
        \caption{4 Pursuers vs. 2 Evaders GCN - Hot Starts heat map}
        \Description{Heat map showing movement patterns with hot start strategies for 4 pursuers vs. 2 evaders.}
    \end{subfigure} \hfill
    \begin{subfigure}[t]{0.45\linewidth}
        \centering
        \includegraphics[height=4cm]{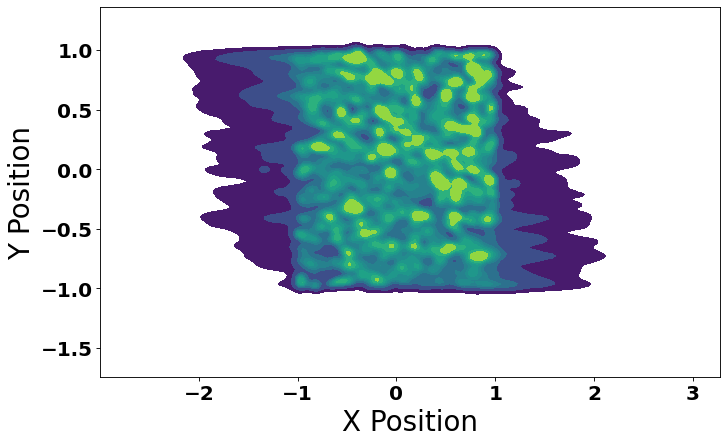}
        \caption{4 Pursuers vs. 2 Evaders Random Start heat map}
        \Description{Heat map showing movement patterns with random start strategies for 4 pursuers vs. 2 evaders.}
    \end{subfigure}
    \caption{Example Heat Maps for Many Pursuers vs. Many Evaders Games. The brighter the color, the more activity in that area of movement. 
 See Appendix section G.4 for heatmaps for all Many Pursuers vs. Many Evaders Games.}
    \label{fig:your_figure_label_2}
\end{figure*}

For One Evader vs. Many Pursuers games, containment by calculating the area of the convex-hull formed by pursuers, checking if the evader was inside the hull using Delaunay triangulation was additionally measured. Further, conditional density plots were used to visualize this.

\begin{figure*}[t]
    \centering
    
    \begin{subfigure}[t]{0.43\linewidth}
        \centering
        \includegraphics[width=\linewidth]{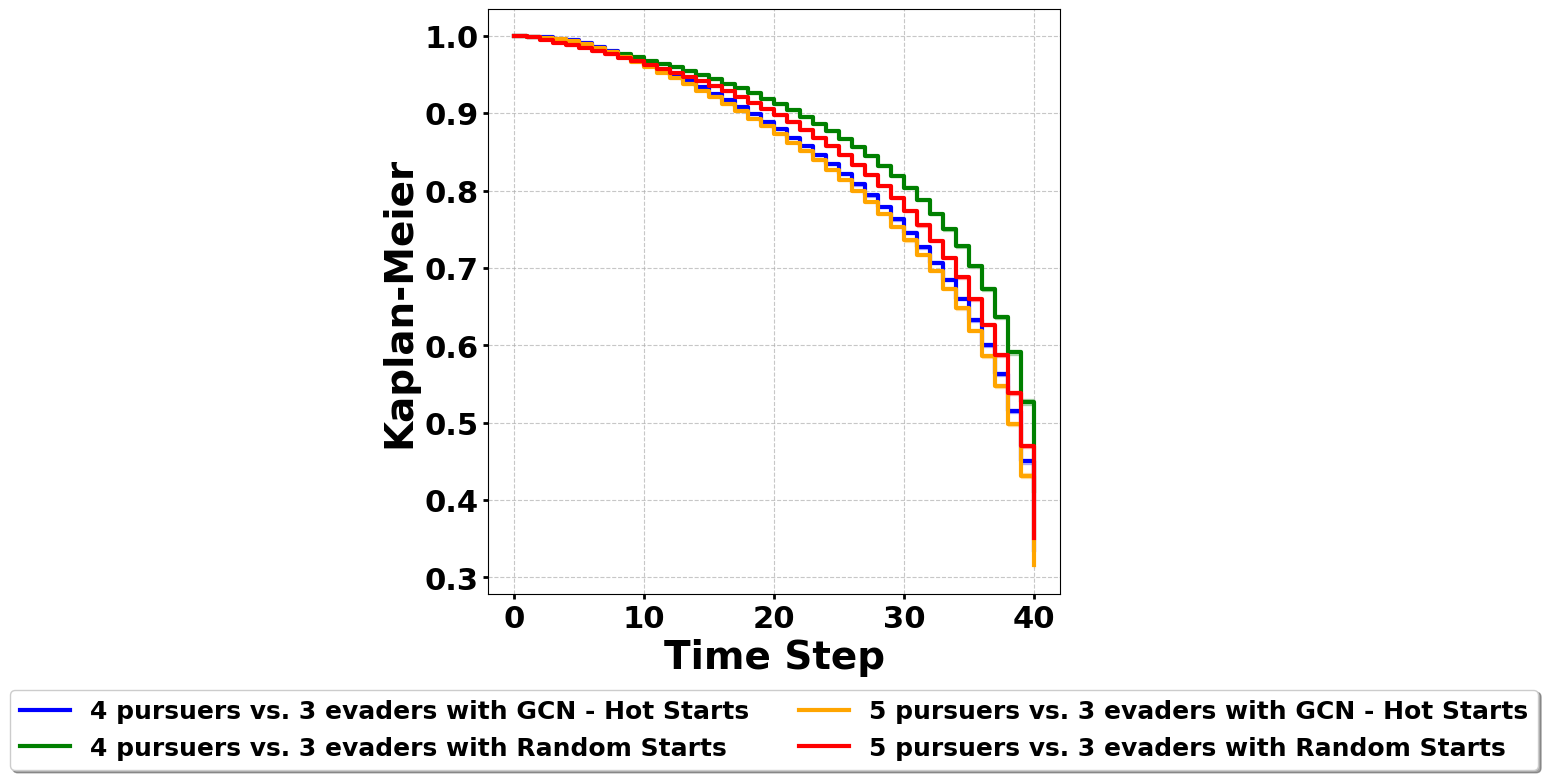}
        \caption{Survival Analysis of Game Simulations with 3 Evaders}
        \Description{Survival analysis plots showing the survival rates of simulations with 3 evaders.}
    \end{subfigure}%
    \hspace{0.04\linewidth}%
    \begin{subfigure}[t]{0.43\linewidth}
        \centering
        \includegraphics[width=\linewidth]{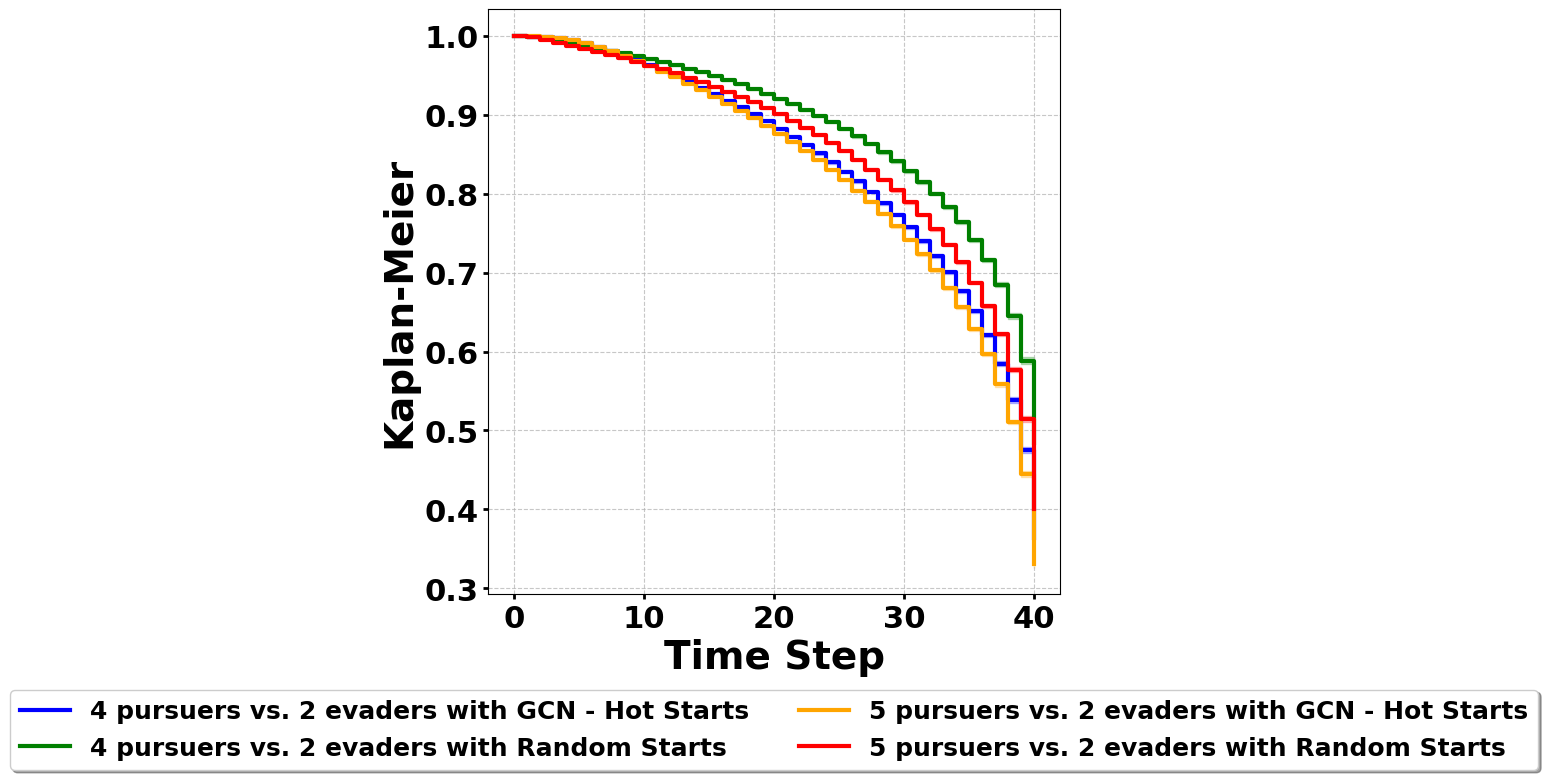}
        \caption{Survival Analysis of Game Simulations with 2 Evaders}
        \Description{Survival analysis plots showing the survival rates of simulations with 2 evaders.}
    \end{subfigure}
    
    \caption{Survival Analysis for Many Pursuers vs. Many Evaders Game Simulations. The x-axis is the time step and the y-axis is the value of the Kaplan-Meier Estimator.}
    \label{fig:survival_analysis}
\end{figure*}

\subsubsection{Many Evaders vs. Many Pursuers Results}
Across all game types, hot starts reduced evader survival rates compared to random starts (see Appendix section G for full results).  In just about every game configuration, the survival rate of evaders for strategies aided by hot starts dropped at a faster rate when compared to without them. Heatmaps revealed distinct movement patterns and activity areas for hot starts, mainly in the activeness of movement and area. Log-rank tests confirmed the statistical significance of our survival analysis, with, $\text{p-value} < .0001$ for all configurations.

\begin{figure*}[t]
    \centering
    \begin{subfigure}[t]{0.45\linewidth}
        \centering
        \includegraphics[height=4cm]{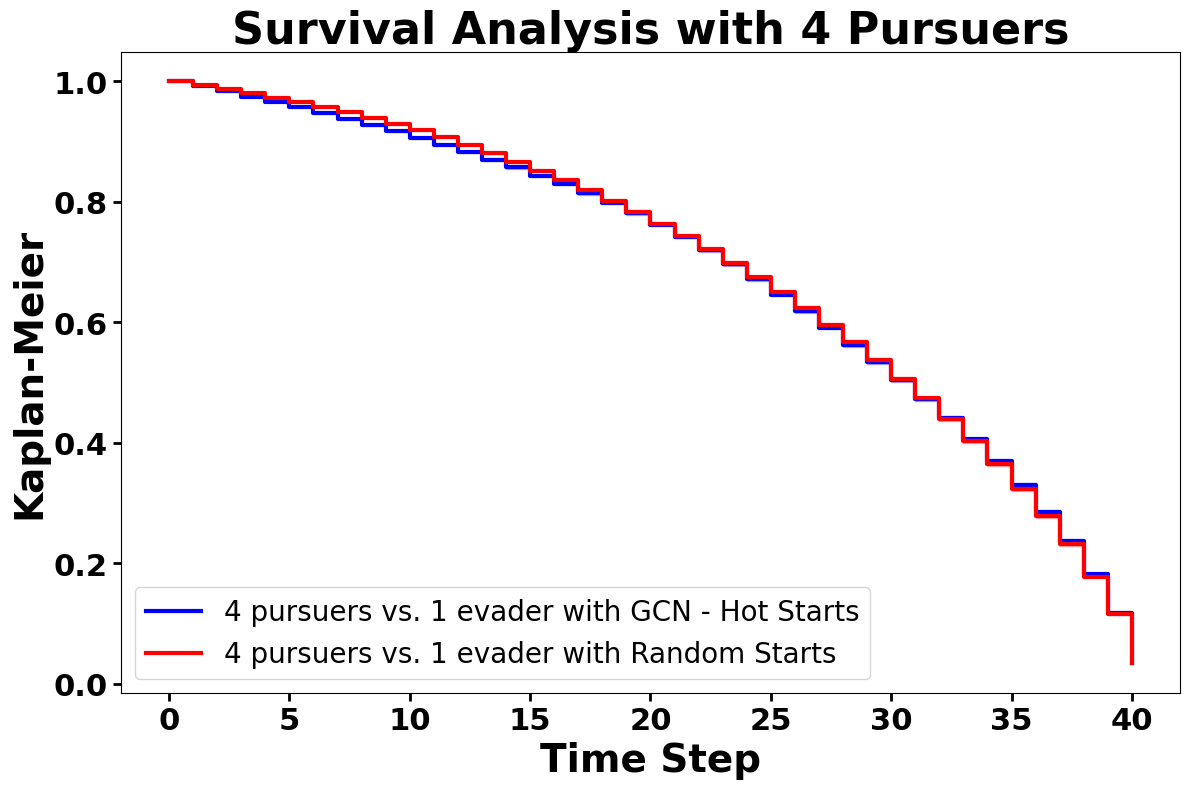}
        \caption{4 Pursuers vs. 1 Evader Survival Analysis}
        \Description{Survival analysis plot showing the survival rates of 1 evader against 5 pursuers.}
    \end{subfigure} \hfill
    \begin{subfigure}[t]{0.45\linewidth}
        \centering
        \includegraphics[height=4cm]{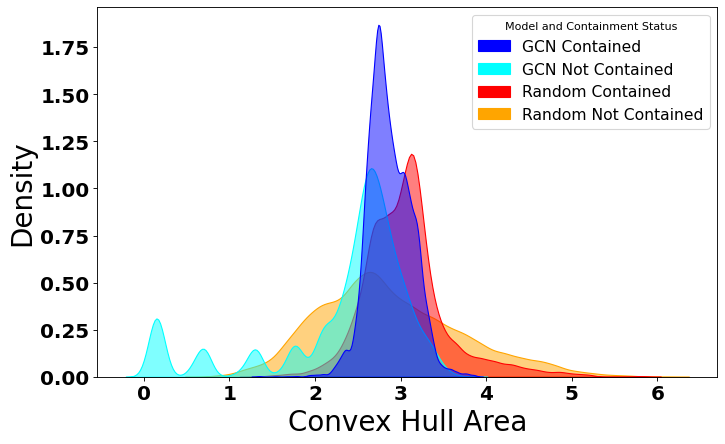}
        \caption{4 Pursuers vs. 1 Evader Conditional Density plot of containment}
        \Description{Conditional Density Plot showing containment metrics for 4 pursuers against 1 evader.}
    \end{subfigure} \\
    \vspace{0.5cm} 
    \begin{subfigure}[t]{0.45\linewidth}
        \centering
        \includegraphics[height=4cm]{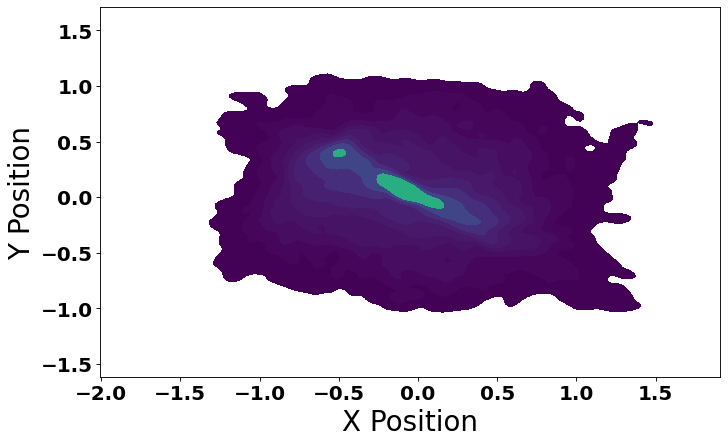}
        \caption{4 Pursuers vs. 1 Evader Hot Start heat map}
        \Description{Heat map showing movement patterns with hot start strategies for 4 pursuers vs. 1 evader.}
    \end{subfigure} \hfill
    \begin{subfigure}[t]{0.45\linewidth}
        \centering
        \includegraphics[height=4cm]{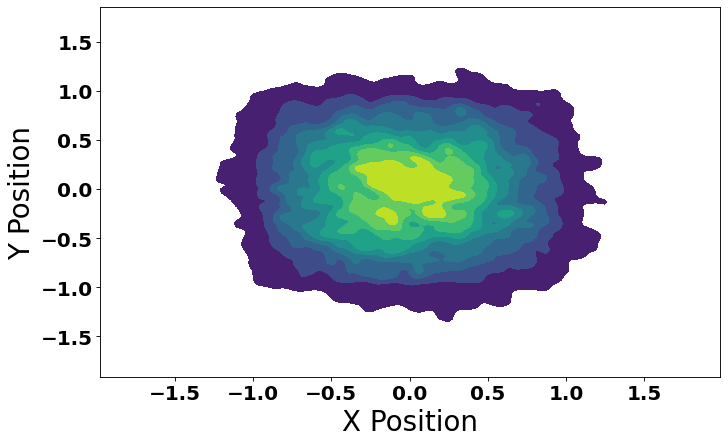}
        \caption{4 Pursuers vs. 1 Evader Random Start heat map}
        \Description{Heat map showing movement patterns with random start strategies for 4 pursuers vs. 1 evader.}
    \end{subfigure} 
    
    \caption{Survival Analysis and movement analysis for 4 pursuers vs. 1 evader games.  See Appendix section G.1 and G.2 for movement analysis and survival analysis tables for all 1 Evader vs. Many Pursuers games.}
    \label{fig:survival_analysis_1_evader}
\end{figure*}

\subsubsection{One Evader vs. Many Pursuers Results}
For One Evader vs. Many Pursuers games, the difference in survival rates between hot starts and random starts was less pronounced (see Appendix Section G for full results). Log-rank tests showed statistically significant differences, except in the 3 pursuers vs. 1 evader game. The heatmaps revealed similar movement patterns across starts, with differences mainly in movement frequency.

Several factors, including the diminishing returns of hot starts in high-pursuer-density scenarios, contribute to this outcome. The high density ensures quick evader interception, reducing the impact of initial configurations; essentially, since the ratio of pursuers to evaders is very high, an advantageous start has less of an impact on movement and the outcomes of the game.

\subsection{Discussion}
\subsubsection{Many Evaders vs. Many Pursuers}
Hot starts provided a clear advantage, as evader survival rates consistently dropped faster compared to random starts, as seen in figures 5a-b (see Table~\ref{tab:survival_probabilities_comparison} for a direct example of 4 Pursuers vs. 2 Evaders games). The statistical significance of these results, supported by log-rank tests with p-values near zero, indicates that hot starts substantially influence game outcomes, favoring pursuers. This highlights the efficacy of our method in capturing and leveraging the spatio-temporal dynamics of agents, allowing pursuers to gain a strategic upper hand.

Moreover, heatmaps revealed that hot starts naturally led to strategic partitioning of the environment, such as dividing the space into distinct regions (e.g., diagonal lines, crescent shapes, or "S" curves, see figure 1 for an example). This partitioning was not pre-programmed but emerged as the GCN learned latent configurations that strategically positioned pursuers from the outset. These emergent patterns ensured that pursuers were never entirely out of position, even in scenarios where random evader placement might suggest no clear advantage. The strategic placement resulted in enhanced capture rates, demonstrating the power of the GCN in discovering and generating effective configurations that maximize pursuit efficiency over many games. 

\subsubsection{One Evader vs. Many Pursuers}
The survival analysis and heatmaps for one evader versus many pursuers differed notably from the many vs. many games. Although the visual differences in survival rates between hot starts and random starts were minimal, log-rank tests still indicated statistically significant results (p-value < 0.05) in all but one game scenario (in 3 pursuers vs. 1 evader games there was no statistical significance). This suggests that, while hot starts have a statistically measurable impact on evader survival, the practical effect is less pronounced due to the overwhelming number of pursuers, which rapidly neutralizes the advantage. High pursuer density ensures that evaders are intercepted quickly, regardless of initial configuration, leading to similar containment outcomes in both hot and random starts. 

\subsubsection{Future Work and Limitations}
This study serves as a proof of concept, establishing the foundational framework for our approach. Future work will concentrate on scaling the network, exploring additional game formats, expanding to larger environments, and applying our methodology to domains such as networked economics, robot swarms, and population initialization. Additionally, we are investigating the incorporation of Generative Adversarial Networks (GANs), Temporal Graph Networks, much more expressive GFS features, and human-in-the-loop systems similar to TacticAI.

While our evaluation spanned a diverse range of game types, limitations remain primarily due to the simplified environments chosen for public accessibility. As a result, the generalizability of hot starts to more complex or varied game scenarios is still an open question. Additionally, the current implementation of the Graph Feature Space relies on heuristics that, while effective for initial studies, lack the sophistication required for capturing all nuances of the underlying dynamics. Future work will focus on integrating the costs derived from the Hamilton-Jacobi-Isaacs (HJI) framework into the objective space, which we expect to yield a more robust performance metric. Furthermore, exploring the impact of hot starts on computing open-loop equilibria represents a promising theoretical direction, potentially offering deeper insights into the dynamics of pursuit-evasion games.

\section{Conclusion} 
This paper presents a novel framework that addresses the problem of strategically placing pursuers when evader locations are unknown, using "hot starts." Hot starts are generated by applying differential game theory algorithms, tracking pursuer movements, and modeling configurations through graphs. By solving a multi-objective optimization problem to identify Pareto-optimal graphs, then proceed to train a GCN on them, and use the GCN to generate hot starts. Empirical evaluations demonstrate that these hot starts offer advantages in terms of faster evader capture, improved containment, and optimized agent movement. Statistical analysis, including log-rank tests, confirms the significance of these results. While this work serves as a proof of concept, it provides a promising foundation for future research.

\begin{acks}
\texttt{This work was generously supported by the Air Force Research Laboratory and the Hap Arnold Scholars Fellowship. Special thanks to Dr. Alex Von Moll and Dr. Isaac Weintraub for their guidance as well.}
\end{acks}



\bibliographystyle{ACM-Reference-Format} 
\bibliography{AAMAS_2025_sample}


\end{document}


\maketitle 


\appendix

\section{Preliminaries}
In this section, we provide some background on differential games and how they fit into the broader sphere of game theory and optimal control theory. The goal is to give readers with some knowledge of game theory or control theory an understanding of differential games using well-known and widely studied concepts. This section is adapted from \textit{A Few Words About Differential Games} \cite{vonmoll2024}.

\subsection{What is a Differential Game?}
Games arise whenever two or more agents interact, each with their own goals. Typically, these goals are encoded by an objective cost function in static games (e.g., a simple utility function) or a cost functional in dynamic games. When agents operate within a system expressed through differential equations, it becomes a differential game. The field of differential game theory encompasses solution concepts, methodologies, applications, and specific instances of differential games, with these concepts often existing in different forms in other areas of control and game theory \cite{isaacs1999differential}.

\subsection{Equilibrium Concepts}

\subsubsection{Nash Equilibrium}
John Nash, along with John von Neumann and others, introduced the concept of an equilibrium, broadly defined as a set of strategies where, holding all other agents' strategies constant, no agent can improve their outcome. This is famously known as the Nash equilibrium:
\begin{equation}
    J(u^*, v^*) \leq J(u, v^*) \quad \forall u \in \mathbb{U}
\end{equation}
where $J$ is a cost function, $u$ is the ego agent's strategy, and $v^*$ is the equilibrium strategy. $\mathbb{U}$ represents the set of possible strategies for the ego agent.

\subsubsection{Other Equilibria}
Nash's equilibrium, originally described for static games, extends to differential games, albeit with increased computational complexity. The solution for a differential game must be tailored to the specific information structure used in the problem formulation. For example, an open-loop Nash equilibrium can exist where players decide on specific control strategies \textit{ex ante} and then "close their eyes" and let the game play out, without updating their strategies or reacting to changes. A more common and practical concept is the state-feedback Nash equilibrium, where player strategies can be a function of the current state. Interestingly, Isaacs \cite{isaacs1999differential} argued that it's not necessary to be in an equilibrium strategy at every moment. Other concepts, such as Stackelberg equilibrium, involve a \textit{leader} who announces their strategy to \textit{followers}, who then respond. The leader’s strategy minimizes their own cost, given the followers' responses. These concepts generally apply to perfect information games, where all players have complete knowledge of the game state (though not necessarily of each other's control policies).

\subsection{Zero-Sum Games}
A zero-sum game is a type of constant-sum game where the sum of the outcomes of the players' strategies always equals zero. In these games, the Nash equilibrium is a saddle-point. For example, consider a cost functional $J$, where the $u$-agent seeks to minimize and the $v$-player seeks to maximize. The saddle-point equilibrium pair of strategies $(u^*, v^*)$ satisfies:
\begin{equation}
    J(u^*, v)  \leq J(u^*, v^*) \leq J(u, v^*) \quad \forall u \in \mathbb{U}, \forall v \in \mathbb{V}
\end{equation}
In this case, the strategy $u^*$ is robust to any possible $v$ that could be used, and vice versa. This is a stronger condition than in general-sum Nash equilibrium, where multiple players could defect and alter the outcome unfavorably.

\subsection{Connecting to Control}
In optimal control, there is typically a single agent with a single cost functional, meaning there are no other decision-makers or control signals in the system. The agent simply seeks to minimize or maximize their cost. One could think of optimal control as a specific type of differential game with one player, or conversely, view differential game theory as optimal control with more than one agent in the system.

\subsection{Linear-Quadratic Games}
When a game is described by a system of linear dynamics, such as:
\begin{equation}
    \dot{\textbf{x}} = A\textbf{x}+ B^u\textbf{u} + B^v\textbf{v}
\end{equation}
with a quadratic cost functional:
\begin{equation}
    J(\textbf{x}, \textbf{u}, \textbf{v}) = \textbf{x}_f^\top Q_f \textbf{x}_f + \int_0^{t_f} \left( \textbf{x}^\top Q_\textbf{x} \textbf{x} + \textbf{u}^\top R^u \textbf{u} - \textbf{v}^\top R^v \textbf{v} \right) dt
\end{equation}
Here, $Q_f$, $Q_\textbf{x}$, $R^u$, and $R^v$ are positive semi-definite cost matrices. This is known as an LQ game. LQ games satisfy Isaacs' key condition for two-player differential games:
\begin{equation}
\min_{\textbf{u}} \max_{\textbf{v}}  J(\textbf{x}, \textbf{u}, \textbf{v}) = \max_{\textbf{v}} \min_{\textbf{u}} J(\textbf{x}, \textbf{u}, \textbf{v})
\end{equation}
This holds because the two agents' controls are separable in the cost functional. This structure is well-established and often allows for analytical solutions, frequently connecting to the Riccati equations in LQ control problems. Researchers often use these games as initial base cases or testing grounds for new ideas.

\subsection{Solutions to Differential Games}
In differential games, we define the solution as a triple $(V(\textbf{x}), \textbf{u}^*, \textbf{v}^*)$, where $V(\textbf{x})$ is the value function, and $\textbf{u}^*$ and $\textbf{v}^*$ are the control strategies for the agents. The value function represents the equilibrium value of the cost functional when the game starts from some initial conditions. For $V$ to be valid, it must be continuously differentiable (a condition we will touch on later) and satisfy the well-known Hamilton-Jacobi-Isaacs (HJI) partial differential equation:
\begin{equation}
    \min_{\textbf{u}} \max_{\textbf{v}} \left[ l(\textbf{x}, \textbf{u}, \textbf{v}, t) + \frac{\partial V}{\partial t} + \nabla_{\textbf{x}} V \cdot \textbf{f}(\textbf{x}, \textbf{u}, \textbf{v}, t) \right] = 0
\end{equation}
where $l$ is the running cost and $\textbf{f} = \dot{\textbf{x}}$. Solving the HJI equation directly is typically infeasible, so alternative solution concepts are often employed.

\subsubsection{Isaacs' Verification Theorem}
In Isaacs' key works, he mentions that it suffices to check whether a candidate value function satisfies the HJI. If it does, then it is a solution to the differential game. One approach often taken by scientists and mathematicians is to guess a strategy for $\textbf{u}$ and $\textbf{v}$ (akin to selecting heading angles and velocities as in our Hot Starts paper), substitute these into $V$, and then substitute $V$ into the HJI.

\subsection{Method of Characteristics}
The Method of Characteristics (MoC) is akin to the indirect method of optimal control, relying on the backward integration of the equilibrium state and costate dynamics from the terminal state set. The Hamiltonian of the system is defined as:
\begin{equation}
    \mathbb{H} = \boldsymbol{\lambda} \cdot \textbf{f} + l 
\end{equation}
where $\boldsymbol{\lambda}$ is the vector of costates, and $l$ is the running cost. From this, we can derive the equilibrium control as a function of costate variables with the necessary conditions for equilibrium:
\begin{equation}
    \textbf{u}^* = \min_{\textbf{u}} \mathbb{H}, \quad \textbf{v}^* = \max_{\textbf{v}} \mathbb{H}
\end{equation}
Ultimately, the terminal costate values can be expressed as a function of the terminal position, allowing for the backward integration of the system's state and costate to capture an equilibrium trajectory (analogous to backward induction in solving a Stackelberg game). This process is repeated over the terminal manifold, essentially filling the space with equilibrium trajectories.

\subsection{Singularities in the Environment}
Although our work did not address this (as we had no obstacles in our environment), we felt it was important to include a section on singularities. Earlier, we mentioned a caveat regarding the regularity requirement of value functions. It is possible that there may be one or more discontinuous surfaces in the environment, or they may not be continuously differentiable. Their presence poses a problem for MoC. If, for a particular terminal state, one were to backward integrate, any trajectory that continues beyond the singularity is incorrect (no longer part of the equilibrium trajectory). This is a significant area of research.

\section{Control Algorithms for Agent Movement}

\subsection{Notation}
\begin{itemize}
    \item $P_i$: Pursuer i
    \item $E_j$: Evader j in many vs. many games
    \item $E$: Evader in one vs. many games
    \item $(x_E, y_E)$: Position of the evader
    \item $\{(x_{P_i}, y_{P_i})\}$: Positions of the pursuers
    \item $\{(x_{E_j}, y_{E_j})\}$: Positions of the evaders in many vs. many games
    \item $v_E$: Velocity of the evader (constant .5)
    \item $v_P$: Velocity of the pursuers (constant 1)
    \item $k$: current time step
    \item $\Delta t$: Change in Time step
    \item $\rho$: Capture radius
    \item $\gamma$: Speed ratio ($\gamma = \frac{v_E}{v_P}$)
    \item $\phi_i$, $\phi_j$: Angles formed by the line connecting the evader to $P_i$ and $P_j$
    \item $\lambda_i$, $\lambda_j$: Line-of-sight angles from pursuers to the evader
    \item $\theta_{ij}$: Weakest link angle between two pursuers $P_i$ and $P_j$
    \item $\Psi_{Es}$, $\Psi_{Ec}$: Functions of distances and angles to the pursuers
    \item $\psi_E$: Optimal heading angle for the evader
    \item $d$: Distance between pursuer $P_i$ and evader
    \item $\theta_{\text{optimal}}$: Optimal heading angle for pursuer $P_i$
    \item $\theta_p$: Optimal heading angle for the pursuer
    \item $\lambda_x$, $\lambda_y$: Co-states corresponding to the x and y coordinates
    \item $H_p$: Hamiltonian of the pursuer
    \item $H_E$: Hamiltonian of the evader
    \item $V$: Value function
\end{itemize}

\subsection{One Evader vs. Many Pursuers}

\subsubsection{Evader Heuristic Control Algorithm}

\paragraph{High-Level Description:}
The evader identifies the weakest link between pursuers by computing the overlapping angle between each pair of pursuers. The pair with the smallest overlapping angle is deemed the weakest link. The evader then computes the optimal heading angle to exploit this weakest link and escape.

\begin{algorithm}[H]
\small
\caption{Evader Control Algorithm for One Evader vs. Many Pursuers Game (Discrete Time)}
\label{alg:evader_policy}
\begin{algorithmic}[1]
\State \textbf{Input:} $(x_E[k], y_E[k])$, $\{(x_{P_i}[k], y_{P_i}[k])\}$, $v_E$, $v_P$, $\Delta t$, $\rho$, $\gamma$
\State \textbf{Output:} $(x_E[k+1], y_E[k+1])$
\State Initialize $\text{min\_theta} \gets \infty$
\State Initialize $\text{weakest\_pair} \gets \text{None}$
\For{each pair of pursuers $(P_i, P_j)$}
    \State Calculate $d_i[k] = \sqrt{(x_{P_i}[k] - x_E[k])^2 + (y_{P_i}[k] - y_E[k])^2}$
    \State Calculate $d_j[k] = \sqrt{(x_{P_j}[k] - x_E[k])^2 + (y_{P_j}[k] - y_E[k])^2}$
    \State Calculate $\phi_i[k] = \begin{cases}
        \arccos\left(\frac{1 - d_i[k]^2}{2d_i[k]}\right) & \text{if } d_i[k]^2 > 1 \\
        0 & \text{otherwise}
    \end{cases}$
    \State Calculate $\phi_j[k] = \begin{cases}
        \arccos\left(\frac{1 - d_j[k]^2}{2d_j[k]}\right) & \text{if } d_j[k]^2 > 1 \\
        0 & \text{otherwise}
    \end{cases}$
    \State Calculate $\lambda_i[k] = \arctan2\left(y_{P_i}[k] - y_E[k],\ x_{P_i}[k] - x_E[k]\right)$
    \State Calculate $\lambda_j[k] = \arctan2\left(y_{P_j}[k] - y_E[k],\ x_{P_j}[k] - x_E[k]\right)$
    \State Calculate $\theta_{ij}[k] = \phi_i[k] + \phi_j[k] - (\lambda_i[k] - \lambda_j[k])$
    \If{$\theta_{ij}[k] < \text{min\_theta}$}
        \State $\text{min\_theta} \gets \theta_{ij}[k]$
        \State $\text{weakest\_pair} \gets (P_i, P_j)$
    \EndIf
\EndFor
\If{$\text{weakest\_pair} \neq \text{None}$}
    \State Let $(P_i, P_j) \gets \text{weakest\_pair}$
    \State Calculate $\Psi_{Es}[k] = \frac{1}{d_j[k]}\left(\cos(\lambda_j[k]) - \frac{\sin(\lambda_j[k])}{\sqrt{\max(d_j[k]^2 - 2,\ 10^{-6})}}\right) - \frac{1}{d_i[k]}\left(\cos(\lambda_i[k]) + \frac{\sin(\lambda_i[k])}{\sqrt{\max(d_i[k]^2 - 2,\ 10^{-6})}}\right)$
    \State Calculate $\Psi_{Ec}[k] = \frac{1}{d_i[k]}\left(\sin(\lambda_i[k]) - \frac{\cos(\lambda_i[k])}{\sqrt{\max(d_i[k]^2 - 2,\ 10^{-6})}}\right) - \frac{1}{d_j[k]}\left(\sin(\lambda_j[k]) + \frac{\cos(\lambda_j[k])}{\sqrt{\max(d_j[k]^2 - 2,\ 10^{-6})}}\right)$
    \State Calculate $\psi_E[k] = \arctan2\left(\Psi_{Es}[k],\ \Psi_{Ec}[k]\right)$

\EndIf
\State Update the evader's position:
\State $x_E[k+1] = x_E[k] + v_E \cos(\psi_E[k]) \Delta t$
\State $y_E[k+1] = y_E[k] + v_E \sin(\psi_E[k]) \Delta t$
\end{algorithmic}
\end{algorithm}

\subsubsection{Pursuers' Control Algorithm}

\paragraph{High-Level Description:}
Pursuers are divided into active and non-active based on their proximity to the evader. Active pursuers compute their optimal heading angle towards the evader while considering the weakest link between pursuers. Non-active pursuers compute the heading angle to a support position to contain the evader.

\begin{algorithm}[H]
\small
\caption{Pursuer Control Algorithm for One Evader vs. Many Pursuers Game (Discrete Time)}
\label{alg:pursuer_policy}
\begin{algorithmic}[1]
\State \textbf{Input:} $(x_E[k], y_E[k])$, $\{(x_{P_i}[k], y_{P_i}[k])\}$, $v_P$, $\Delta t$, $\rho$, $\gamma$
\State \textbf{Output:} $\{(x_{P_i}[k+1], y_{P_i}[k+1])\}$
\For{each pursuer $P_i$}
    \State Calculate $d[k] = \sqrt{(x_E[k] - x_{P_i}[k])^2 + (y_E[k] - y_{P_i}[k])^2}$
    \If{$d[k] \leq \rho$}
        \State $x_{P_i}[k+1] = x_{P_i}[k]$
        \State $y_{P_i}[k+1] = y_{P_i}[k]$
    \Else
        \State $x_{P_i}^{[k]} = x_{P_i}[k] + \frac{\gamma (x_E[k] - x_{P_i}[k]) + \rho}{1 - \gamma^2}$
        \State $y_{P_i}^{[k]} = y_{P_i}[k] + \frac{\gamma (y_E[k] - y_{P_i}[k]) + \rho}{1 - \gamma^2}$
    \EndIf
    \State Calculate $\theta_{\text{optimal}}[k] = \arctan2 \left(y_{P_i}^{[k]} - y_{P_i}[k],\ x_{P_i}^{[k]} - x_{P_i}[k]\right)$
    \State Update the pursuer's position:
    \State $x_{P_i}[k+1] = x_{P_i}[k] + v_P \cos(\theta_{\text{optimal}}[k]) \Delta t$
    \State $y_{P_i}[k+1] = y_{P_i}[k] + v_P \sin(\theta_{\text{optimal}}[k]) \Delta t$
\EndFor 
\end{algorithmic}
\end{algorithm}

\subsection{Many Evaders vs. Many Pursuers}

For many-vs-many scenarios ($\gamma = \frac{1}{2} < 1$), where pursuers are faster than evaders, we employ a hybrid approach combining geometry-based controls with PDE solutions. Specifically, we use a first-order upwind finite difference scheme with directional bias: backward differences are selected for negative gradients and forward differences for positive gradients. This scheme provides numerical stability while respecting the information flow direction in the HJI equation, consistent with methods described in \cite{Mitchell2005TimeDependentHJ, Osher2003LevelSM}. The Hamiltonian computation explicitly accounts for the minimizing pursuer and maximizing evader roles, ensuring appropriate upwind direction selection based on agent type.

\subsubsection{Geometric Measures and Hybrid Control}

Our hybrid approach uses a weighted parameter $\alpha = 0.5$ to combine geometric distance metrics with PDE value function information in the assignment cost matrix for the Hungarian algorithm. The geometric measures specifically include:

\begin{enumerate}
    \item Euclidean distances between pursuers and evaders, used in the assignment cost matrix:
    \begin{equation}
        d_{ij} = \|\mathbf{p}_i - \mathbf{e}_j\|_2
    \end{equation}
    
    \item Normalized direction vectors for determining movement trajectories:
    \begin{equation}
        \hat{\mathbf{d}}_{ij} = \frac{\mathbf{e}_j - \mathbf{p}_i}{\|\mathbf{e}_j - \mathbf{p}_i\|_2}
    \end{equation}
    
    \item Relative bearing angles for optimal interception paths:
    \begin{equation}
        \theta_{ij} = \arctan2(\hat{\mathbf{d}}_{ij}[1], \hat{\mathbf{d}}_{ij}[0])
    \end{equation}
    
    \item Proximity-based avoidance vectors that allow evaders to compute escape directions:
    \begin{equation}
        \mathbf{a}_j = \sum_{i=1}^{N_p} \frac{\mathbf{e}_j - \mathbf{p}_i}{\|\mathbf{e}_j - \mathbf{p}_i\|_2^2} \cdot \mathbf{1}_{\|\mathbf{e}_j - \mathbf{p}_i\| < r_{avoid}}
    \end{equation}
\end{enumerate}

Pursuers are allocated to evaders by minimizing this hybrid cost using the Hungarian algorithm, and control actions are computed by combining gradient information from the PDE with direct geometric vectors:

\begin{equation}
\mathbf{u}_{\text{final}} = \alpha \cdot \mathbf{u}_{\text{geometric}} + (1-\alpha) \cdot \mathbf{u}_{\text{gradient}}
\end{equation}

where $\mathbf{u}_{\text{geometric}}$ is derived from the instantaneous geometric relationship between agents, and $\mathbf{u}_{\text{gradient}}$ comes from the spatial derivatives of the value function.

\subsubsection{Upwind Scheme for HJI Equation}

We compute the upwind derivatives of the value function as follows:

\begin{align}
V_x^{\text{forward}} &= \frac{V(i+1,j) - V(i,j)}{\Delta x} \\
V_x^{\text{backward}} &= \frac{V(i,j) - V(i-1,j)}{\Delta x} \\
V_y^{\text{forward}} &= \frac{V(i,j+1) - V(i,j)}{\Delta y} \\
V_y^{\text{backward}} &= \frac{V(i,j) - V(i,j-1)}{\Delta y}
\end{align}

Then, for pursuers (minimizing player) and evaders (maximizing player), we select the appropriate derivatives:

\begin{align}
V_x^{\text{pursuer}} &= \begin{cases}
    V_x^{\text{backward}} & \text{if } V_x^{\text{forward}} \geq 0 \\
    V_x^{\text{forward}} & \text{otherwise}
\end{cases} \\
V_x^{\text{evader}} &= \begin{cases}
    V_x^{\text{forward}} & \text{if } V_x^{\text{forward}} \geq 0 \\
    V_x^{\text{backward}} & \text{otherwise}
\end{cases}
\end{align}

This directional bias ensures numerical stability while correctly modeling the min-max nature of the game.

\subsubsection{System Dynamics and Value Function Evolution}

The system dynamics for the Many Evaders vs. Many Pursuers game in discrete time are defined as follows. Let there be $N_p$ pursuers and $N_e$ evaders. The state vector $\mathbf{x}[k]$ at time step $k$ comprises the positions of all pursuers and evaders:

\[
\mathbf{x}[k] = \begin{bmatrix}
x_{P_1}[k] \\
y_{P_1}[k] \\
\vdots \\
x_{P_{N_p}}[k] \\
y_{P_{N_p}}[k] \\
x_{E_1}[k] \\
y_{E_1}[k] \\
\vdots \\
x_{E_{N_e}}[k] \\
y_{E_{N_e}}[k] \\
\end{bmatrix}
\]

The discrete-time system dynamics $\mathbf{x}[k+1]$ are given by:

\[
\mathbf{x}[k+1] = \mathbf{x}[k] + \Delta t \cdot \mathbf{f}(\mathbf{x}[k], \mathbf{u}_p[k], \mathbf{u}_E[k])
\]

where the control inputs $\mathbf{u}_p[k]$ for pursuers and $\mathbf{u}_E[k]$ for evaders are derived from our hybrid approach.

The discrete-time HJI equation is formulated as:

\[
V[k+1] = V[k] - \Delta t \cdot H(\mathbf{x}[k], \nabla V[k])
\]

where $H$ is the Hamiltonian:

\[
H(\mathbf{x}, \nabla V) = \min_{\mathbf{u}_p} \max_{\mathbf{u}_E} \nabla V \cdot \mathbf{f}(\mathbf{x}, \mathbf{u}_p, \mathbf{u}_E)
\]

Evaders employ escape strategies based on gradient ascent of the value function combined with geometric avoidance of nearby pursuers, computing avoidance vectors that are inversely proportional to the squared distance from approaching pursuers when they enter a predefined avoidance radius.
\section{Environment}
The environment in use was the PettingZoo Simple Tag MPE Environment. This is a predator prey environment, it's a discrete time step environment and is a continuous space environment. We used the environment without obstacles, and augmented the environment to our need; we also ignored rewards related to collisions, as those were not needed for our situation.  With the use of open source tools, we aim to make our code available to the public once it's fully gone through the release process. 
\vspace{10cm}
\section{Training Results}
\begin{figure}[h!]
    \centering
    \begin{minipage}[b]{0.9\linewidth}
        \centering
        \includegraphics[width=\linewidth]{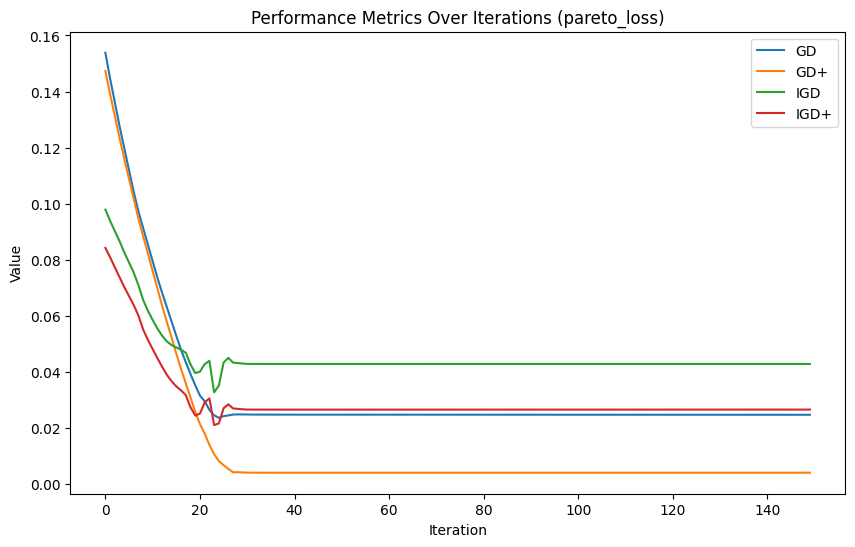}
        \caption{Generational Difference (GD), Generational Difference Plus (GD+), Inverse Generational Difference (IGD), and Inverse Generational Difference Plus (IGD+).}
        \label{subfig:image2}
    \end{minipage}
    \hspace{0.01\linewidth}  
    \begin{minipage}[b]{0.9\linewidth}
        \centering
    \includegraphics[width=\linewidth]{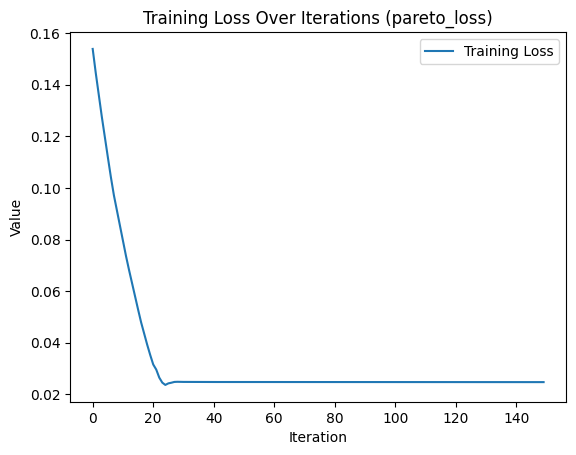}
        \caption{Pareto Loss during GCN training.}
        \label{subfig:image3}
    \end{minipage}
    \caption{Training metrics for GCN: GD, GD+, IGD, IGD+, and Pareto Loss.}
    \label{fig:collective}
\end{figure}

In this section, we include our training results for a batch size of 1024, learning rate of $1e^{-5}$, Adam optimizer, weight decay of $1e^{-2}$, and we trained for 150 epochs. We trained our network on an L4 GPU, which took about 2 hours. Over training, we measured GD, GD+, IGD, and IGD+ \cite{pymoo} to give us an idea of how the generated graphs relate to our Pareto values from our GCS. Generational Difference (GD) and Generational Difference Plus (GD+), can be expressed as:
\begin{equation}
    GD = \left( \frac{1}{|P|} \sum_{x \in P} \left( d(x, P^*) \right)^p \right)^{1/p}
\end{equation}
\begin{equation}
    GD^+ = \left( \frac{1}{|P|} \sum_{x \in P} \left( d^+(x, P^*) \right)^p \right)^{1/p}
\end{equation}
where $P$ is the set of approximate Pareto-optimal solutions, $P^*$ is the set of true Pareto-optimal solutions, $d(x, P^*)$ is the minimum Euclidean distance from solution $x$ to the true Pareto front $P^*$, $d^+(x, P^*)$ is the perpendicular distance from $x$ to $P^*$, and $p$ is usually set to 2.

Inverse Generational Difference (IGD) and Inverse Generational Difference Plus (IGD+) are expressed as:
\begin{equation}
    IGD = \left( \frac{1}{|P^*|} \sum_{x^* \in P^*} \left( d(x^*, P) \right)^p \right)^{1/p}
\end{equation}
\begin{equation}
    IGD^+ = \left( \frac{1}{|P^*|} \sum_{x^* \in P^*} \left( d^+(x^*, P) \right)^p \right)^{1/p}
\end{equation}
where $d(x^*, P)$ is the minimum Euclidean distance from the true Pareto-optimal solution $x^*$ to the approximate Pareto front $P$.

Finally, Hypervolume (HV) is expressed as:
\begin{equation} 
    HV = \text{Lebesgue Measure} \left( \bigcup_{x \in P} [f_1(x), r_1] \times [f_2(x), r_2] \times \cdots \times [f_m(x), r_m] \right)
\end{equation} 
where $r = (r_1, r_2, \dots, r_m)$ is a reference point dominated by all Pareto-optimal solutions, and $f_i(x)$ is the $i$-th objective function value of solution $x$.
\vspace{5cm}
\section{Mathematical Examples of $ U_{\text{heading}}$ and $U_{\text{capture}}$ }
To illustrate the correctness and behavior of \( U_{\text{capture}} \) and \( U_{\text{heading}} \), consider the following examples:

\paragraph{Example 1: \( U_{\text{capture}} \)}
\begin{itemize}
    \item Scenario 1: \( d_i = r \)
    \begin{align*}
    U_{\text{capture}} &= 2 \left(1 - \frac{1}{1 + e^{\frac{-r + r}{r}}}\right) \\
    &= 2 \left(1 - \frac{1}{1 + e^{0}}\right) \\
    &= 2 \left(1 - \frac{1}{2}\right) \\
    &= 2 \times 0.5 \\
    &= 1
    \end{align*}
    \textbf{Interpretation}: When the pursuer is exactly at the capture radius distance from the evader, the capture potential \( U_{\text{capture}} \) is 1.

    \item Scenario 2: \( d_i < r \) (e.g., \( d_i = 0 \))
    \begin{align*}
    U_{\text{capture}} &= 2 \left(1 - \frac{1}{1 + e^{\frac{-0 + r}{r}}}\right) \\
    &= 2 \left(1 - \frac{1}{1 + e^{1}}\right) \\
    &= 2 \left(1 - \frac{1}{1 + 2.71828}\right) \\
    &= 2 \left(1 - \frac{1}{3.71828}\right) \\
    &= 2 \times 0.73206 \\
    &= 1.46412
    \end{align*}
    \textbf{Interpretation}: When the pursuer is at the same position as the evader (\( d_i = 0 \)), the capture potential \( U_{\text{capture}} \) exceeds 1, indicating a higher likelihood of capture.

    \item Scenario 3: \( d_i > r \) (e.g., \( d_i = 2r \))
    \begin{align*}
    U_{\text{capture}} &= 2 \left(1 - \frac{1}{1 + e^{\frac{-2r + r}{r}}}\right) \\
    &= 2 \left(1 - \frac{1}{1 + e^{-1}}\right) \\
    &= 2 \left(1 - \frac{1}{1 + 0.36788}\right) \\
    &= 2 \left(1 - \frac{1}{1.36788}\right) \\
    &= 2 \times 0.26768 \\
    &= 0.53536
    \end{align*}
    \textbf{Interpretation}: When the pursuer is twice the capture radius distance away from the evader, the capture potential \( U_{\text{capture}} \) decreases to approximately 0.535, indicating a lower likelihood of capture.
\end{itemize}

\paragraph{Example 2: \( U_{\text{heading}} \)}
\begin{itemize}
    \item Scenario 1: Perfect Alignment (\( \theta = 0^\circ \))
    \begin{align*}
    \mathbf{v}_p &= \begin{bmatrix} 1 \\ 0 \end{bmatrix} \quad \text{(Velocity vector pointing along the positive x-axis)} \\
    \mathbf{e}_1 - \mathbf{p} &= \begin{bmatrix} 1 \\ 0 \end{bmatrix} \quad \text{(Evader is directly ahead)} \\
    \theta &= \arccos \left( \frac{\mathbf{v}_p \cdot (\mathbf{e}_1 - \mathbf{p})}{\| \mathbf{v}_p \| \| \mathbf{e}_1 - \mathbf{p} \| + \epsilon} \right) \\
    &= \arccos \left( \frac{1 \times 1 + 0 \times 0}{1 \times 1 + 10^{-9}} \right) \\
    &\approx \arccos(1) \\
    &\approx 0^\circ \\
    U_{\text{heading}} & \approx \frac{1}{1} \times 0^\circ \\
    & \approx 0^\circ
    \end{align*}
    \textbf{Interpretation}: When the pursuer's heading is perfectly aligned with the direction to the evader, \( U_{\text{heading}} \) is 0, indicating optimal alignment for direct interception.

    \item Scenario 2: Partial Alignment (\( \theta = 60^\circ \))
    \begin{align*}
    \mathbf{v}_p &= \begin{bmatrix} \cos(60^\circ) \\ \sin(60^\circ) \end{bmatrix} = \begin{bmatrix} 0.5 \\ 0.8660 \end{bmatrix} \\
    \mathbf{e}_1 - \mathbf{p} &= \begin{bmatrix} 1 \\ 0 \end{bmatrix} \\
    \theta &= \arccos \left( \frac{0.5 \times 1 + 0.8660 \times 0}{\sqrt{0.5^2 + 0.8660^2} \times \sqrt{1^2 + 0^2} + \epsilon} \right) \\
    &= \arccos \left( \frac{0.5}{1 + 10^{-9}} \right) \\
    &\approx \arccos(0.5) \\
    &\approx 60^\circ \\
    U_{\text{heading}} &\approx \frac{1}{1} \times 60^\circ \\
    &\approx 60^\circ
    \end{align*}
    \textbf{Interpretation}: When the pursuer's heading forms a \( 60^\circ \) angle with the direction to the evader, \( U_{\text{heading}} \) is \( 60^\circ \), indicating a moderate alignment.

    \item Scenario 3: Orthogonal Alignment (\( \theta = 90^\circ \))
    \begin{align*}
    \mathbf{v}_p &= \begin{bmatrix} 0 \\ 1 \end{bmatrix} \quad \text{(Velocity vector pointing along the positive y-axis)} \\
    \mathbf{e}_1 - \mathbf{p} &= \begin{bmatrix} 1 \\ 0 \end{bmatrix} \\
    \theta &= \arccos \left( \frac{0 \times 1 + 1 \times 0}{\sqrt{0^2 + 1^2} \times \sqrt{1^2 + 0^2} + \epsilon} \right) \\
    &= \arccos(0) \\
    &= 90^\circ \\
    U_{\text{heading}} &= \frac{1}{1} \times 90^\circ \\
    &= 90^\circ
    \end{align*}
    \textbf{Interpretation}: When the pursuer's heading is orthogonal to the direction to the evader, \( U_{\text{heading}} \) is \( 90^\circ \), indicating no alignment.
    
    \item Scenario 4: Opposite Alignment (\( \theta = 180^\circ \))
    \begin{align*}
    \mathbf{v}_p &= \begin{bmatrix} -1 \\ 0 \end{bmatrix} \quad \text{(Velocity vector pointing along the negative x-axis)} \\
    \mathbf{e}_1 - \mathbf{p} &= \begin{bmatrix} 1 \\ 0 \end{bmatrix} \\
    \theta &= \arccos \left( \frac{-1 \times 1 + 0 \times 0}{\sqrt{(-1)^2 + 0^2} \times \sqrt{1^2 + 0^2} + \epsilon} \right) \\
    &\approx \arccos(-1) \\
    &\approx 180^\circ \\
    U_{\text{heading}} &\approx \frac{1}{1} \times 180^\circ \\
    &\approx 180^\circ
    \end{align*}
    \textbf{Interpretation}: When the pursuer's heading is directly opposite to the direction to the evader, \( U_{\text{heading}} \) is \( 180^\circ \), indicating maximum misalignment.
\end{itemize}

\section{Kaplan-Meier Estimator}
\label{appendix:kaplan-meier}

In our work, the Kaplan-Meier estimator is used to estimate the survival probability of an evader over the course of the game. The Kaplan-Meier estimator is commonly used to estimate the survival probability of cancer patients or to determine how long a piece of machinery will last \cite{rich2010practical, kaplan1958nonparametric}. The formula is expressed as:

\begin{equation}
    \hat{S}(t) = \prod_{t_i \leq t} \left(1 - \frac{d_i}{n_i}\right)
    \label{eq:kaplan-meier}
\end{equation}

where:
\begin{itemize}
    \item \( t_i \) are the ordered times at which events (evader captures) occur.
    \item \( d_i \) is the number of evaders captured at time \( t_i \).
    \item \( n_i \) is the number of evaders at risk just before time \( t_i \) (essentially the evader still alive).
\end{itemize}

\subsection{Application in Our Study}

The Kaplan-Meier estimator provides a step-wise estimate of the survival probability \( \hat{S}(t) \), accounting for censored data—evaders that are not captured within the observation period. By applying this estimator to our simulation data, we were able to:

\begin{itemize}
    \item \textbf{Estimate Evader Survival Probabilities:} Calculate the probability that an evader remains uncaptured up to each time point \( t \).
    \item \textbf{Compare Pursuer Configurations:} Analyze and compare the effectiveness of different pursuer configurations (e.g., hot starts vs. random starts) based on the estimated survival probabilities.
    \item \textbf{Assess Statistical Significance:} Utilize the Kaplan-Meier survival curves in conjunction with log-rank tests to determine the statistical significance of differences observed between strategies.
\end{itemize}

This approach allowed us to rigorously evaluate the performance improvements introduced by our pursuer configuration strategies, ensuring that the observed benefits are both meaningful and statistically validated.

\section{Additional Plots and Tables}
In this section, we include further plots and tables on movement in simulations. 4 Pursuer vs. 2 Evader games results, are given in the paper, but here we include more to further our point. 

\subsection{Many vs. Many Supplementary Tables}
We see across the board the impact of hot starts on survival probability as in each case, by simply using a hot start we are able to reduce the survival probability of evaders across each game format tested. 

\begin{table*}[htbp]
    \centering
    \caption{Survival Probabilities for 5 Pursuers vs. 2 Evaders}
    \label{tab:survival_p5v2}
    
    \begin{minipage}[t]{0.48\textwidth}
        \centering
        \caption*{5 Pursuers vs. 2 Evaders with GCN - Hot Starts}
        \begin{tabular}{@{}cc@{}}
            \toprule
            \textbf{Time} & \textbf{Survival Probability} \\ \midrule
            0    & 1.000000 \\
            1    & 0.999222 \\
            2    & 0.998300 \\
            3    & 0.996939 \\
            4    & 0.994560 \\
            5    & 0.990941 \\
            6    & 0.986354 \\
            7    & 0.980970 \\
            8    & 0.974989 \\
            9    & 0.968652 \\
            10   & 0.961855 \\
            11   & 0.954719 \\
            12   & 0.947278 \\
            13   & 0.939514 \\
            14   & 0.931382 \\
            15   & 0.922949 \\
            16   & 0.914204 \\
            17   & 0.905176 \\
            18   & 0.895805 \\
            19   & 0.886094 \\
            20   & 0.875959 \\
            21   & 0.865351 \\
            22   & 0.854188 \\
            23   & 0.842500 \\
            24   & 0.830273 \\
            25   & 0.817456 \\
            26   & 0.803995 \\
            27   & 0.789788 \\
            28   & 0.774721 \\
            29   & 0.758697 \\
            30   & 0.741516 \\
            31   & 0.723015 \\
            32   & 0.702835 \\
            33   & 0.680740 \\
            34   & 0.656146 \\
            35   & 0.628434 \\
            36   & 0.596409 \\
            37   & 0.558314 \\
            38   & 0.510671 \\
            39   & 0.445101 \\
            40   & 0.330532 \\ \bottomrule
        \end{tabular}
    \end{minipage}%
    \hfill
    \begin{minipage}[t]{0.48\textwidth}
        \centering
        \caption*{5 Pursuers vs. 2 Evaders with Random Starts}
        \begin{tabular}{@{}cc@{}}
            \toprule
            \textbf{Time} & \textbf{Survival Probability} \\ \midrule
            0    & 1.000000 \\
            1    & 0.998020 \\
            2    & 0.994668 \\
            3    & 0.991215 \\
            4    & 0.987615 \\
            5    & 0.983908 \\
            6    & 0.980029 \\
            7    & 0.975901 \\
            8    & 0.971578 \\
            9    & 0.967084 \\
            10   & 0.962386 \\
            11   & 0.957494 \\
            12   & 0.952343 \\
            13   & 0.946922 \\
            14   & 0.941247 \\
            15   & 0.935277 \\
            16   & 0.929030 \\
            17   & 0.922492 \\
            18   & 0.915537 \\
            19   & 0.908179 \\
            20   & 0.900417 \\
            21   & 0.892182 \\
            22   & 0.883387 \\
            23   & 0.874053 \\
            24   & 0.864181 \\
            25   & 0.853671 \\
            26   & 0.842459 \\
            27   & 0.830490 \\
            28   & 0.817720 \\
            29   & 0.804057 \\
            30   & 0.789255 \\
            31   & 0.773004 \\
            32   & 0.755122 \\
            33   & 0.735234 \\
            34   & 0.712862 \\
            35   & 0.687413 \\
            36   & 0.657744 \\
            37   & 0.622012 \\
            38   & 0.576916 \\
            39   & 0.513946 \\
            40   & 0.400981 \\ \bottomrule
        \end{tabular}
    \end{minipage}
\end{table*}

\vspace{10pt} 

\begin{table*}[htbp]
    \centering
    \caption{Survival Probabilities for 4 Pursuers vs. 3 Evaders}
    \label{tab:survival_p4v3}
    
    \begin{minipage}[t]{0.48\textwidth}
        \centering
        \caption*{4 Pursuers vs. 3 Evaders with GCN - Hot Starts}
        \begin{tabular}{@{}cc@{}}
            \toprule
            \textbf{Time} & \textbf{Survival Probability} \\ \midrule
            0    & 1.000000 \\
            1    & 0.999092 \\
            2    & 0.998024 \\
            3    & 0.996518 \\
            4    & 0.994117 \\
            5    & 0.990603 \\
            6    & 0.986178 \\
            7    & 0.980965 \\
            8    & 0.975210 \\
            9    & 0.969067 \\
            10   & 0.962570 \\
            11   & 0.955685 \\
            12   & 0.948512 \\
            13   & 0.941056 \\
            14   & 0.933264 \\
            15   & 0.925127 \\
            16   & 0.916647 \\
            17   & 0.907805 \\
            18   & 0.898566 \\
            19   & 0.888968 \\
            20   & 0.878970 \\
            21   & 0.868551 \\
            22   & 0.857614 \\
            23   & 0.846139 \\
            24   & 0.834086 \\
            25   & 0.821397 \\
            26   & 0.807958 \\
            27   & 0.793756 \\
            28   & 0.778695 \\
            29   & 0.762651 \\
            30   & 0.745433 \\
            31   & 0.726878 \\
            32   & 0.706741 \\
            33   & 0.684626 \\
            34   & 0.660110 \\
            35   & 0.632504 \\
            36   & 0.600679 \\
            37   & 0.562786 \\
            38   & 0.515387 \\
            39   & 0.450105 \\
            40   & 0.335590 \\ \bottomrule
        \end{tabular}
    \end{minipage}%
    \hfill
    \begin{minipage}[t]{0.48\textwidth}
        \centering
        \caption*{4 Pursuers vs. 3 Evaders with Random Starts}
        \begin{tabular}{@{}cc@{}}
            \toprule
            \textbf{Time} & \textbf{Survival Probability} \\ \midrule
            0    & 1.000000 \\
            1    & 0.998310 \\
            2    & 0.995495 \\
            3    & 0.992602 \\
            4    & 0.989600 \\
            5    & 0.986452 \\
            6    & 0.983164 \\
            7    & 0.979680 \\
            8    & 0.976016 \\
            9    & 0.972165 \\
            10   & 0.968093 \\
            11   & 0.963801 \\
            12   & 0.959254 \\
            13   & 0.954455 \\
            14   & 0.949379 \\
            15   & 0.943999 \\
            16   & 0.938291 \\
            17   & 0.932209 \\
            18   & 0.925741 \\
            19   & 0.918875 \\
            20   & 0.911560 \\
            21   & 0.903789 \\
            22   & 0.895473 \\
            23   & 0.886663 \\
            24   & 0.877266 \\
            25   & 0.867232 \\
            26   & 0.856392 \\
            27   & 0.844800 \\
            28   & 0.832285 \\
            29   & 0.818680 \\
            30   & 0.803850 \\
            31   & 0.787633 \\
            32   & 0.769758 \\
            33   & 0.749969 \\
            34   & 0.727702 \\
            35   & 0.702192 \\
            36   & 0.672407 \\
            37   & 0.636503 \\
            38   & 0.590763 \\
            39   & 0.526715 \\
            40   & 0.411715 \\ \bottomrule
        \end{tabular}
    \end{minipage}
\end{table*}

\vspace{10pt} 

\begin{table*}[htbp]
    \centering
    \caption{Survival Probabilities for 5 Pursuers vs. 3 Evaders}
    \label{tab:survival_p5v3}
    
    \begin{minipage}[t]{0.48\textwidth}
        \centering
        \caption*{5 Pursuers vs. 3 Evaders with GCN - Hot Starts}
        \begin{tabular}{@{}cc@{}}
            \toprule
            \textbf{Time} & \textbf{Survival Probability} \\ \midrule
            0    & 1.000000 \\
            1    & 0.998993 \\
            2    & 0.997815 \\
            3    & 0.996199 \\
            4    & 0.993568 \\
            5    & 0.989759 \\
            6    & 0.984922 \\
            7    & 0.979366 \\
            8    & 0.973246 \\
            9    & 0.966679 \\
            10   & 0.959752 \\
            11   & 0.952511 \\
            12   & 0.944972 \\
            13   & 0.937122 \\
            14   & 0.928993 \\
            15   & 0.920544 \\
            16   & 0.911727 \\
            17   & 0.902554 \\
            18   & 0.893002 \\
            19   & 0.883076 \\
            20   & 0.872743 \\
            21   & 0.861976 \\
            22   & 0.850758 \\
            23   & 0.838999 \\
            24   & 0.826674 \\
            25   & 0.813681 \\
            26   & 0.799921 \\
            27   & 0.785332 \\
            28   & 0.769855 \\
            29   & 0.753342 \\
            30   & 0.735631 \\
            31   & 0.716578 \\
            32   & 0.695787 \\
            33   & 0.672959 \\
            34   & 0.647618 \\
            35   & 0.619072 \\
            36   & 0.586154 \\
            37   & 0.547038 \\
            38   & 0.498303 \\
            39   & 0.431647 \\
            40   & 0.316023 \\ \bottomrule
        \end{tabular}
    \end{minipage}%
    \hfill
    \begin{minipage}[t]{0.48\textwidth}
        \centering
        \caption*{5 Pursuers vs. 3 Evaders with Random Starts}
        \begin{tabular}{@{}cc@{}}
            \toprule
            \textbf{Time} & \textbf{Survival Probability} \\ \midrule
            0    & 1.000000 \\
            1    & 0.997978 \\
            2    & 0.994643 \\
            3    & 0.991232 \\
            4    & 0.987706 \\
            5    & 0.983977 \\
            6    & 0.980093 \\
            7    & 0.975994 \\
            8    & 0.971676 \\
            9    & 0.967206 \\
            10   & 0.962549 \\
            11   & 0.957687 \\
            12   & 0.952562 \\
            13   & 0.947123 \\
            14   & 0.941317 \\
            15   & 0.935106 \\
            16   & 0.928461 \\
            17   & 0.921335 \\
            18   & 0.913756 \\
            19   & 0.905696 \\
            20   & 0.897185 \\
            21   & 0.888120 \\
            22   & 0.878488 \\
            23   & 0.868210 \\
            24   & 0.857298 \\
            25   & 0.845724 \\
            26   & 0.833380 \\
            27   & 0.820070 \\
            28   & 0.805687 \\
            29   & 0.790146 \\
            30   & 0.773343 \\
            31   & 0.755066 \\
            32   & 0.734925 \\
            33   & 0.712675 \\
            34   & 0.687779 \\
            35   & 0.659481 \\
            36   & 0.626621 \\
            37   & 0.587363 \\
            38   & 0.538064 \\
            39   & 0.469927 \\
            40   & 0.350472 \\ \bottomrule
        \end{tabular}
    \end{minipage}
\end{table*}

\vspace{10pt}

\begin{table*}[htbp]
    \centering
    \caption{Survival Probabilities for 4 Pursuers vs. 4 Evaders}
    \label{tab:survival_p4v4}
    
    \begin{minipage}[t]{0.48\textwidth}
        \centering
        \caption*{4 Pursuers vs. 4 Evaders with GCN - Hot Starts}
        \begin{tabular}{@{}cc@{}}
            \toprule
            \textbf{Time} & \textbf{Survival Probability} \\ \midrule
            0    & 1.000000 \\
            1    & 0.999123 \\
            2    & 0.998092 \\
            3    & 0.996676 \\
            4    & 0.994383 \\
            5    & 0.991085 \\
            6    & 0.986928 \\
            7    & 0.982059 \\
            8    & 0.976650 \\
            9    & 0.970822 \\
            10   & 0.964563 \\
            11   & 0.958040 \\
            12   & 0.951187 \\
            13   & 0.944028 \\
            14   & 0.936554 \\
            15   & 0.928742 \\
            16   & 0.920618 \\
            17   & 0.912157 \\
            18   & 0.903338 \\
            19   & 0.894107 \\
            20   & 0.884480 \\
            21   & 0.874427 \\
            22   & 0.863919 \\
            23   & 0.852904 \\
            24   & 0.841324 \\
            25   & 0.829138 \\
            26   & 0.816269 \\
            27   & 0.802641 \\
            28   & 0.788166 \\
            29   & 0.772719 \\
            30   & 0.756101 \\
            31   & 0.738148 \\
            32   & 0.718582 \\
            33   & 0.697008 \\
            34   & 0.672980 \\
            35   & 0.645808 \\
            36   & 0.614503 \\
            37   & 0.577210 \\
            38   & 0.530444 \\
            39   & 0.465796 \\
            40   & 0.351763 \\ \bottomrule
        \end{tabular}
    \end{minipage}%
    \hfill
    \begin{minipage}[t]{0.48\textwidth}
        \centering
        \caption*{4 Pursuers vs. 4 Evaders with Random Starts}
        \begin{tabular}{@{}cc@{}}
            \toprule
            \textbf{Time} & \textbf{Survival Probability} \\ \midrule
            0    & 1.000000 \\
            1    & 0.998219 \\
            2    & 0.995346 \\
            3    & 0.992389 \\
            4    & 0.989340 \\
            5    & 0.986164 \\
            6    & 0.982834 \\
            7    & 0.979360 \\
            8    & 0.975713 \\
            9    & 0.971886 \\
            10   & 0.967864 \\
            11   & 0.963565 \\
            12   & 0.958984 \\
            13   & 0.954065 \\
            14   & 0.948824 \\
            15   & 0.943268 \\
            16   & 0.937344 \\
            17   & 0.931068 \\
            18   & 0.924379 \\
            19   & 0.917233 \\
            20   & 0.909538 \\
            21   & 0.901298 \\
            22   & 0.892398 \\
            23   & 0.882879 \\
            24   & 0.872709 \\
            25   & 0.861794 \\
            26   & 0.850149 \\
            27   & 0.837632 \\
            28   & 0.824169 \\
            29   & 0.809639 \\
            30   & 0.793828 \\
            31   & 0.776627 \\
            32   & 0.757696 \\
            33   & 0.736540 \\
            34   & 0.712833 \\
            35   & 0.685960 \\
            36   & 0.654861 \\
            37   & 0.617421 \\
            38   & 0.569983 \\
            39   & 0.503936 \\
            40   & 0.386834 \\ \bottomrule
        \end{tabular}
    \end{minipage}
\end{table*}

\vspace{10pt} 

\begin{table*}[htbp]
    \centering
    \caption{Survival Probabilities for 5 Pursuers vs. 4 Evaders}
    \label{tab:survival_p5v4}
    
    \begin{minipage}[t]{0.48\textwidth}
        \centering
        \caption*{5 Pursuers vs. 4 Evaders with GCN - Hot Starts}
        \begin{tabular}{@{}cc@{}}
            \toprule
            \textbf{Time} & \textbf{Survival Probability} \\ \midrule
            0    & 1.000000 \\
            1    & 0.999006 \\
            2    & 0.997873 \\
            3    & 0.996323 \\
            4    & 0.993675 \\
            5    & 0.989947 \\
            6    & 0.985225 \\
            7    & 0.979768 \\
            8    & 0.973911 \\
            9    & 0.967684 \\
            10   & 0.961106 \\
            11   & 0.954195 \\
            12   & 0.946975 \\
            13   & 0.939429 \\
            14   & 0.931536 \\
            15   & 0.923328 \\
            16   & 0.914780 \\
            17   & 0.905861 \\
            18   & 0.896497 \\
            19   & 0.886741 \\
            20   & 0.876601 \\
            21   & 0.866038 \\
            22   & 0.854959 \\
            23   & 0.843363 \\
            24   & 0.831203 \\
            25   & 0.818387 \\
            26   & 0.804829 \\
            27   & 0.790460 \\
            28   & 0.775195 \\
            29   & 0.758881 \\
            30   & 0.741385 \\
            31   & 0.722554 \\
            32   & 0.702146 \\
            33   & 0.679747 \\
            34   & 0.654883 \\
            35   & 0.626821 \\
            36   & 0.594471 \\
            37   & 0.556039 \\
            38   & 0.507923 \\
            39   & 0.441791 \\
            40   & 0.326174 \\ \bottomrule
        \end{tabular}
    \end{minipage}%
    \hfill
    \begin{minipage}[t]{0.48\textwidth}
        \centering
        \caption*{5 Pursuers vs. 4 Evaders with Random Starts}
        \begin{tabular}{@{}cc@{}}
            \toprule
            \textbf{Time} & \textbf{Survival Probability} \\ \midrule
            0    & 1.000000 \\
            1    & 0.997989 \\
            2    & 0.994624 \\
            3    & 0.991179 \\
            4    & 0.987612 \\
            5    & 0.983902 \\
            6    & 0.980035 \\
            7    & 0.975990 \\
            8    & 0.971731 \\
            9    & 0.967220 \\
            10   & 0.962384 \\
            11   & 0.957233 \\
            12   & 0.951731 \\
            13   & 0.945845 \\
            14   & 0.939566 \\
            15   & 0.932819 \\
            16   & 0.925681 \\
            17   & 0.918108 \\
            18   & 0.910039 \\
            19   & 0.901412 \\
            20   & 0.892220 \\
            21   & 0.882472 \\
            22   & 0.872068 \\
            23   & 0.861005 \\
            24   & 0.849204 \\
            25   & 0.836610 \\
            26   & 0.823179 \\
            27   & 0.808853 \\
            28   & 0.793553 \\
            29   & 0.777094 \\
            30   & 0.759270 \\
            31   & 0.739848 \\
            32   & 0.718577 \\
            33   & 0.695026 \\
            34   & 0.668694 \\
            35   & 0.638804 \\
            36   & 0.604321 \\
            37   & 0.563265 \\
            38   & 0.511989 \\
            39   & 0.441693 \\
            40   & 0.319653 \\ \bottomrule
        \end{tabular}
    \end{minipage}
\end{table*}

\vspace{10pt} 

\begin{table*}[htbp]
    \centering
    \caption{Survival Probabilities for 4 Pursuers vs. 5 Evaders}
    \label{tab:survival_p4v5}
    
    \begin{minipage}[t]{0.48\textwidth}
        \centering
        \caption*{4 Pursuers vs. 5 Evaders with GCN - Hot Starts}
        \begin{tabular}{@{}cc@{}}
            \toprule
            \textbf{Time} & \textbf{Survival Probability} \\ \midrule
            0    & 1.000000 \\
            1    & 0.999111 \\
            2    & 0.998097 \\
            3    & 0.996702 \\
            4    & 0.994433 \\
            5    & 0.991136 \\
            6    & 0.986983 \\
            7    & 0.982107 \\
            8    & 0.976650 \\
            9    & 0.971018 \\
            10   & 0.964981 \\
            11   & 0.958622 \\
            12   & 0.951959 \\
            13   & 0.945030 \\
            14   & 0.937834 \\
            15   & 0.930331 \\
            16   & 0.922515 \\
            17   & 0.914364 \\
            18   & 0.905882 \\
            19   & 0.897037 \\
            20   & 0.887789 \\
            21   & 0.878097 \\
            22   & 0.867943 \\
            23   & 0.857282 \\
            24   & 0.846027 \\
            25   & 0.834158 \\
            26   & 0.821579 \\
            27   & 0.808211 \\
            28   & 0.793962 \\
            29   & 0.778708 \\
            30   & 0.762245 \\
            31   & 0.744428 \\
            32   & 0.724961 \\
            33   & 0.703507 \\
            34   & 0.679653 \\
            35   & 0.652665 \\
            36   & 0.621402 \\
            37   & 0.584048 \\
            38   & 0.537081 \\
            39   & 0.472121 \\
            40   & 0.357325 \\ \bottomrule
        \end{tabular}
    \end{minipage}%
    \hfill
    \begin{minipage}[t]{0.48\textwidth}
        \centering
        \caption*{4 Pursuers vs. 5 Evaders with Random Starts}
        \begin{tabular}{@{}cc@{}}
            \toprule
            \textbf{Time} & \textbf{Survival Probability} \\ \midrule
            0    & 1.000000 \\
            1    & 0.998219 \\
            2    & 0.995346 \\
            3    & 0.992389 \\
            4    & 0.989340 \\
            5    & 0.986164 \\
            6    & 0.982834 \\
            7    & 0.979360 \\
            8    & 0.975713 \\
            9    & 0.971886 \\
            10   & 0.967864 \\
            11   & 0.963565 \\
            12   & 0.958984 \\
            13   & 0.954065 \\
            14   & 0.948824 \\
            15   & 0.943268 \\
            16   & 0.937344 \\
            17   & 0.931068 \\
            18   & 0.924379 \\
            19   & 0.917233 \\
            20   & 0.909538 \\
            21   & 0.901298 \\
            22   & 0.892398 \\
            23   & 0.882879 \\
            24   & 0.872709 \\
            25   & 0.861794 \\
            26   & 0.850149 \\
            27   & 0.837632 \\
            28   & 0.824169 \\
            29   & 0.809639 \\
            30   & 0.793828 \\
            31   & 0.776627 \\
            32   & 0.757696 \\
            33   & 0.736540 \\
            34   & 0.712833 \\
            35   & 0.685960 \\
            36   & 0.654861 \\
            37   & 0.617386 \\
            38   & 0.570022 \\
            39   & 0.503928 \\
            40   & 0.386160 \\ \bottomrule
        \end{tabular}
    \end{minipage}
\end{table*}

\vspace{10pt} 

\begin{table*}[htbp]
    \centering
    \caption{Survival Probabilities for 5 Pursuers vs. 5 Evaders}
    \label{tab:survival_p5v5}
    
    \begin{minipage}[t]{0.48\textwidth}
        \centering
        \caption*{5 Pursuers vs. 5 Evaders with GCN - Hot Starts}
        \begin{tabular}{@{}cc@{}}
            \toprule
            \textbf{Time} & \textbf{Survival Probability} \\ \midrule
            0    & 1.000000 \\
            1    & 0.999070 \\
            2    & 0.998004 \\
            3    & 0.996513 \\
            4    & 0.993849 \\
            5    & 0.990070 \\
            6    & 0.985211 \\
            7    & 0.979584 \\
            8    & 0.973390 \\
            9    & 0.966780 \\
            10   & 0.959835 \\
            11   & 0.952516 \\
            12   & 0.944874 \\
            13   & 0.936892 \\
            14   & 0.928621 \\
            15   & 0.920041 \\
            16   & 0.911140 \\
            17   & 0.901895 \\
            18   & 0.892243 \\
            19   & 0.882160 \\
            20   & 0.871633 \\
            21   & 0.860602 \\
            22   & 0.849035 \\
            23   & 0.836887 \\
            24   & 0.824146 \\
            25   & 0.810707 \\
            26   & 0.796522 \\
            27   & 0.781482 \\
            28   & 0.765492 \\
            29   & 0.748426 \\
            30   & 0.730154 \\
            31   & 0.710437 \\
            32   & 0.689038 \\
            33   & 0.665604 \\
            34   & 0.639634 \\
            35   & 0.610420 \\
            36   & 0.576881 \\
            37   & 0.537168 \\
            38   & 0.487778 \\
            39   & 0.420328 \\
            40   & 0.303443 \\ \bottomrule
        \end{tabular}
    \end{minipage}%
    \hfill
    \begin{minipage}[t]{0.48\textwidth}
        \centering
        \caption*{5 Pursuers vs. 5 Evaders with Random Starts}
        \begin{tabular}{@{}cc@{}}
            \toprule
            \textbf{Time} & \textbf{Survival Probability} \\ \midrule
            0    & 1.000000 \\
            1    & 0.997945 \\
            2    & 0.994521 \\
            3    & 0.990998 \\
            4    & 0.987299 \\
            5    & 0.983443 \\
            6    & 0.979421 \\
            7    & 0.975179 \\
            8    & 0.970706 \\
            9    & 0.965971 \\
            10   & 0.960984 \\
            11   & 0.955716 \\
            12   & 0.950118 \\
            13   & 0.944165 \\
            14   & 0.937817 \\
            15   & 0.931047 \\
            16   & 0.923794 \\
            17   & 0.916031 \\
            18   & 0.907752 \\
            19   & 0.898889 \\
            20   & 0.889436 \\
            21   & 0.879392 \\
            22   & 0.868607 \\
            23   & 0.857125 \\
            24   & 0.844874 \\
            25   & 0.831829 \\
            26   & 0.817994 \\
            27   & 0.803284 \\
            28   & 0.787488 \\
            29   & 0.770572 \\
            30   & 0.752280 \\
            31   & 0.732417 \\
            32   & 0.710725 \\
            33   & 0.686770 \\
            34   & 0.659997 \\
            35   & 0.629638 \\
            36   & 0.594463 \\
            37   & 0.552714 \\
            38   & 0.500678 \\
            39   & 0.429612 \\
            40   & 0.307052 \\ \bottomrule
        \end{tabular}
    \end{minipage}
\end{table*}
\FloatBarrier
\subsection{One vs. Many Supplementary Tables}
In this subsection, we see the survival probabilities for the one vs. many games.  We see that hot starts had a much smaller impact here, this is because the pursuers have such a large number advantage in comparison to the evader. The advantage given by a hot start has little influence on the evader survival probabilities, as already the pursuers have a large advantage. 
\begin{table*}[htbp]
    \centering
    \caption{Survival Probabilities for 2 Pursuers vs. 1 Evader}
    \label{tab:survival_p2v1}
    \begin{minipage}[t]{0.48\textwidth}
        \centering
        \caption*{2 Pursuers vs. 1 Evader with GCN - Hot Starts}
        \begin{tabular}{@{}cc@{}}
            \toprule
            \textbf{Time} & \textbf{Survival Probability} \\ \midrule
            0    & 1.000000 \\
            1    & 0.991712 \\
            2    & 0.983283 \\
            3    & 0.974511 \\
            4    & 0.965517 \\
            5    & 0.956183 \\
            6    & 0.946949 \\
            7    & 0.937354 \\
            8    & 0.927597 \\
            9    & 0.917843 \\
            10   & 0.907954 \\
            11   & 0.898163 \\
            12   & 0.888222 \\
            13   & 0.878467 \\
            14   & 0.868690 \\
            15   & 0.859034 \\
            16   & 0.849344 \\
            17   & 0.839081 \\
            18   & 0.827845 \\
            19   & 0.814938 \\
            20   & 0.799280 \\
            21   & 0.781656 \\
            22   & 0.761806 \\
            23   & 0.739184 \\
            24   & 0.714617 \\
            25   & 0.689271 \\
            26   & 0.663882 \\
            27   & 0.639366 \\
            28   & 0.616202 \\
            29   & 0.594326 \\
            30   & 0.573309 \\
            31   & 0.551781 \\
            32   & 0.528637 \\
            33   & 0.502271 \\
            34   & 0.470700 \\
            35   & 0.431475 \\
            36   & 0.385005 \\
            37   & 0.330575 \\
            38   & 0.266333 \\
            39   & 0.191560 \\
            40   & 0.089746 \\ \bottomrule
        \end{tabular}
    \end{minipage}%
    \hfill
    \begin{minipage}[t]{0.48\textwidth}
        \centering
        \caption*{2 Pursuers vs. 1 Evader with Random Starts}
        \begin{tabular}{@{}cc@{}}
            \toprule
            \textbf{Time} & \textbf{Survival Probability} \\ \midrule
            0    & 1.000000 \\
            1    & 0.993463 \\
            2    & 0.986801 \\
            3    & 0.979894 \\
            4    & 0.972637 \\
            5    & 0.965018 \\
            6    & 0.957064 \\
            7    & 0.948788 \\
            8    & 0.940033 \\
            9    & 0.930603 \\
            10   & 0.920607 \\
            11   & 0.910143 \\
            12   & 0.899190 \\
            13   & 0.887629 \\
            14   & 0.875333 \\
            15   & 0.862641 \\
            16   & 0.849322 \\
            17   & 0.835078 \\
            18   & 0.820283 \\
            19   & 0.804828 \\
            20   & 0.788348 \\
            21   & 0.771300 \\
            22   & 0.753276 \\
            23   & 0.733586 \\
            24   & 0.712938 \\
            25   & 0.691772 \\
            26   & 0.669866 \\
            27   & 0.646732 \\
            28   & 0.623076 \\
            29   & 0.597842 \\
            30   & 0.570885 \\
            31   & 0.542512 \\
            32   & 0.512252 \\
            33   & 0.480236 \\
            34   & 0.445796 \\
            35   & 0.407903 \\
            36   & 0.366012 \\
            37   & 0.318201 \\
            38   & 0.263206 \\
            39   & 0.195693 \\
            40   & 0.094324 \\ \bottomrule
        \end{tabular}
    \end{minipage}
\end{table*}
\vspace{10pt}
\begin{table*}[htbp]
    \centering
    \caption{Survival Probabilities for 3 Pursuers vs. 1 Evader}
    \label{tab:survival_p3v1}
    
    \begin{minipage}[t]{0.48\textwidth}
        \centering
        \caption*{3 Pursuers vs. 1 Evader with GCN - Hot Starts}
        \begin{tabular}{@{}cc@{}}
            \toprule
            \textbf{Time} & \textbf{Survival Probability} \\ \midrule
            0    & 1.000000 \\
            1    & 0.991542 \\
            2    & 0.982940 \\
            3    & 0.973990 \\
            4    & 0.964908 \\
            5    & 0.955572 \\
            6    & 0.946125 \\
            7    & 0.936274 \\
            8    & 0.926023 \\
            9    & 0.915364 \\
            10   & 0.904409 \\
            11   & 0.893104 \\
            12   & 0.881432 \\
            13   & 0.869092 \\
            14   & 0.856554 \\
            15   & 0.843397 \\
            16   & 0.829981 \\
            17   & 0.815929 \\
            18   & 0.801278 \\
            19   & 0.785568 \\
            20   & 0.768335 \\
            21   & 0.749677 \\
            22   & 0.729725 \\
            23   & 0.708374 \\
            24   & 0.685581 \\
            25   & 0.661486 \\
            26   & 0.636702 \\
            27   & 0.611265 \\
            28   & 0.585529 \\
            29   & 0.559782 \\
            30   & 0.533947 \\
            31   & 0.507748 \\
            32   & 0.480724 \\
            33   & 0.452001 \\
            34   & 0.421050 \\
            35   & 0.386524 \\
            36   & 0.346016 \\
            37   & 0.298554 \\
            38   & 0.242127 \\
            39   & 0.173000 \\
            40   & 0.073006 \\ \bottomrule
        \end{tabular}
    \end{minipage}%
    \hfill
    \begin{minipage}[t]{0.48\textwidth}
        \centering
        \caption*{3 Pursuers vs. 1 Evader with Random Starts}
        \begin{tabular}{@{}cc@{}}
            \toprule
            \textbf{Time} & \textbf{Survival Probability} \\ \midrule
            0    & 1.000000 \\
            1    & 0.993342 \\
            2    & 0.986558 \\
            3    & 0.979592 \\
            4    & 0.972267 \\
            5    & 0.964462 \\
            6    & 0.956222 \\
            7    & 0.947448 \\
            8    & 0.938145 \\
            9    & 0.928314 \\
            10   & 0.917903 \\
            11   & 0.906899 \\
            12   & 0.895380 \\
            13   & 0.882983 \\
            14   & 0.869967 \\
            15   & 0.856226 \\
            16   & 0.841693 \\
            17   & 0.826461 \\
            18   & 0.810387 \\
            19   & 0.793565 \\
            20   & 0.775779 \\
            21   & 0.757549 \\
            22   & 0.738543 \\
            23   & 0.718767 \\
            24   & 0.698388 \\
            25   & 0.676883 \\
            26   & 0.653929 \\
            27   & 0.629687 \\
            28   & 0.603854 \\
            29   & 0.576597 \\
            30   & 0.548011 \\
            31   & 0.517688 \\
            32   & 0.485342 \\
            33   & 0.451186 \\
            34   & 0.415005 \\
            35   & 0.376779 \\
            36   & 0.335710 \\
            37   & 0.290361 \\
            38   & 0.237999 \\
            39   & 0.173898 \\
            40   & 0.078950 \\ \bottomrule
        \end{tabular}
    \end{minipage}
\end{table*}

\vspace{10pt} 

\begin{table*}[htbp]
    \centering
    \caption{Survival Probabilities for 4 Pursuers vs. 1 Evader}
    \label{tab:survival_p4v1}
    
    \begin{minipage}[t]{0.48\textwidth}
        \centering
        \caption*{4 Pursuers vs. 1 Evader with GCN - Hot Starts}
        \begin{tabular}{@{}cc@{}}
            \toprule
            \textbf{Time} & \textbf{Survival Probability} \\ \midrule
            0    & 1.000000 \\
            1    & 0.991600 \\
            2    & 0.983057 \\
            3    & 0.974306 \\
            4    & 0.965630 \\
            5    & 0.956510 \\
            6    & 0.947177 \\
            7    & 0.937420 \\
            8    & 0.927343 \\
            9    & 0.916736 \\
            10   & 0.905824 \\
            11   & 0.894433 \\
            12   & 0.882366 \\
            13   & 0.869911 \\
            14   & 0.856854 \\
            15   & 0.843144 \\
            16   & 0.828870 \\
            17   & 0.813872 \\
            18   & 0.797834 \\
            19   & 0.780508 \\
            20   & 0.761804 \\
            21   & 0.741454 \\
            22   & 0.719728 \\
            23   & 0.696346 \\
            24   & 0.671677 \\
            25   & 0.645555 \\
            26   & 0.618582 \\
            27   & 0.590867 \\
            28   & 0.562506 \\
            29   & 0.533408 \\
            30   & 0.503428 \\
            31   & 0.472656 \\
            32   & 0.440436 \\
            33   & 0.406234 \\
            34   & 0.369310 \\
            35   & 0.329163 \\
            36   & 0.284956 \\
            37   & 0.236425 \\
            38   & 0.181495 \\
            39   & 0.117632 \\
            40   & 0.034466 \\ \bottomrule
        \end{tabular}
    \end{minipage}%
    \hfill
    \begin{minipage}[t]{0.48\textwidth}
        \centering
        \caption*{4 Pursuers vs. 1 Evader with Random Starts}
        \begin{tabular}{@{}cc@{}}
            \toprule
            \textbf{Time} & \textbf{Survival Probability} \\ \midrule
            0    & 1.000000 \\
            1    & 0.993475 \\
            2    & 0.986826 \\
            3    & 0.979912 \\
            4    & 0.972702 \\
            5    & 0.965028 \\
            6    & 0.956867 \\
            7    & 0.948178 \\
            8    & 0.938947 \\
            9    & 0.928949 \\
            10   & 0.918191 \\
            11   & 0.906515 \\
            12   & 0.895996 \\
            13   & 0.882618 \\
            14   & 0.868438 \\
            15   & 0.851609 \\
            16   & 0.836041 \\
            17   & 0.819477 \\
            18   & 0.801867 \\
            19   & 0.783452 \\
            20   & 0.763779 \\
            21   & 0.742971 \\
            22   & 0.721122 \\
            23   & 0.698146 \\
            24   & 0.674348 \\
            25   & 0.649481 \\
            26   & 0.623199 \\
            27   & 0.595733 \\
            28   & 0.567184 \\
            29   & 0.537419 \\
            30   & 0.506212 \\
            31   & 0.473409 \\
            32   & 0.438982 \\
            33   & 0.402917 \\
            34   & 0.364424 \\
            35   & 0.323153 \\
            36   & 0.291866 \\
            37   & 0.244088 \\
            38   & 0.189949 \\
            39   & 0.125860 \\
            40   & 0.035190 \\ \bottomrule
        \end{tabular}
    \end{minipage}
\end{table*}

\vspace{10pt} 

\begin{table*}[htbp]
    \centering
    \caption{Survival Probabilities for 5 Pursuers vs. 1 Evader}
    \label{tab:survival_p5v1}
    
    \begin{minipage}[t]{0.48\textwidth}
        \centering
        \caption*{5 Pursuers vs. 1 Evader with GCN - Hot Starts}
        \begin{tabular}{@{}cc@{}}
            \toprule
            \textbf{Time} & \textbf{Survival Probability} \\ \midrule
            0    & 1.000000 \\
            1    & 0.991545 \\
            2    & 0.982947 \\
            3    & 0.974100 \\
            4    & 0.965133 \\
            5    & 0.955943 \\
            6    & 0.946558 \\
            7    & 0.936875 \\
            8    & 0.926854 \\
            9    & 0.916600 \\
            10   & 0.905950 \\
            11   & 0.894928 \\
            12   & 0.883454 \\
            13   & 0.871427 \\
            14   & 0.858801 \\
            15   & 0.845529 \\
            16   & 0.831392 \\
            17   & 0.816614 \\
            18   & 0.801055 \\
            19   & 0.784466 \\
            20   & 0.766864 \\
            21   & 0.747930 \\
            22   & 0.727870 \\
            23   & 0.706147 \\
            24   & 0.683110 \\
            25   & 0.658689 \\
            26   & 0.633246 \\
            27   & 0.607021 \\
            28   & 0.579705 \\
            29   & 0.551502 \\
            30   & 0.522533 \\
            31   & 0.492414 \\
            32   & 0.460528 \\
            33   & 0.426760 \\
            34   & 0.390509 \\
            35   & 0.351276 \\
            36   & 0.308013 \\
            37   & 0.259501 \\
            38   & 0.204193 \\
            39   & 0.138545 \\
            40   & 0.049294 \\ \bottomrule
        \end{tabular}
    \end{minipage}%
    \hfill
    \begin{minipage}[t]{0.48\textwidth}
        \centering
        \caption*{5 Pursuers vs. 1 Evader with Random Starts}
        \begin{tabular}{@{}cc@{}}
            \toprule
            \textbf{Time} & \textbf{Survival Probability} \\ \midrule
            0    & 1.000000 \\
            1    & 0.993080 \\
            2    & 0.986032 \\
            3    & 0.978725 \\
            4    & 0.971133 \\
            5    & 0.963078 \\
            6    & 0.954553 \\
            7    & 0.945502 \\
            8    & 0.938145 \\
            9    & 0.928314 \\
            10   & 0.917903 \\
            11   & 0.906899 \\
            12   & 0.895380 \\
            13   & 0.882983 \\
            14   & 0.869967 \\
            15   & 0.856226 \\
            16   & 0.841693 \\
            17   & 0.826461 \\
            18   & 0.810387 \\
            19   & 0.793565 \\
            20   & 0.775779 \\
            21   & 0.757549 \\
            22   & 0.738543 \\
            23   & 0.718767 \\
            24   & 0.698388 \\
            25   & 0.676883 \\
            26   & 0.653929 \\
            27   & 0.629687 \\
            28   & 0.603854 \\
            29   & 0.576597 \\
            30   & 0.548011 \\
            31   & 0.517688 \\
            32   & 0.485342 \\
            33   & 0.451186 \\
            34   & 0.415005 \\
            35   & 0.376779 \\
            36   & 0.335710 \\
            37   & 0.290361 \\
            38   & 0.237999 \\
            39   & 0.173898 \\
            40   & 0.040376 \\ \bottomrule
        \end{tabular}
    \end{minipage}
\end{table*}
\vspace{10pt}
\FloatBarrier
\subsection{One vs. Many Pursuers Movement Analysis}
In this section, we show conditional density plots of containment, and we also show heatmaps of pursuers movement, the brighter the color the more active the pursuers were. 
\subsubsection{1 Evader vs. 2 Pursuers Movement Analysis}
\begin{figure}[H]
    \centering
    \begin{subfigure}[t]{0.45\linewidth}
        \centering
        \includegraphics[height=4cm]{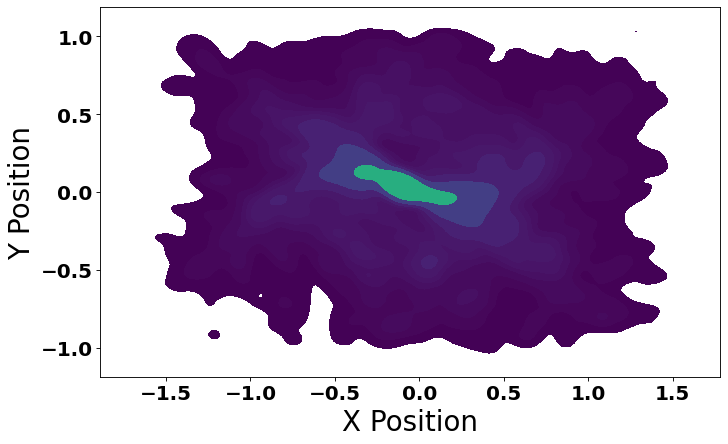}
        \caption{Heat map showing movement patterns with hot start strategies for 2 pursuers vs. 1 evader.}
    \end{subfigure} \hfill
    \begin{subfigure}[t]{0.45\linewidth}
        \centering
        \includegraphics[height=4cm]{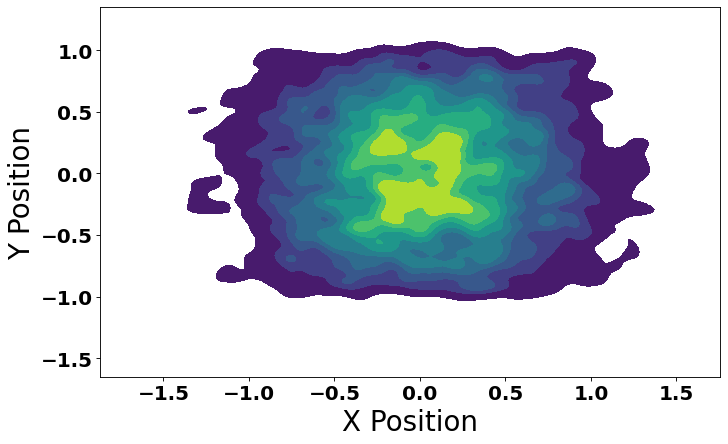}
        \caption{Heat map showing movement patterns with random start strategies for 2 pursuers vs. 1 evader.}
    \end{subfigure}
    \caption{Movement analysis for 2 Pursuers vs. 1 Evader games.}
    \label{fig:survival_analysis_2_pursuers}
\end{figure}

\subsubsection{1 Evader vs. 3 Pursuers Movement Analysis}
\begin{figure}[H]
    \centering
    \begin{subfigure}[t]{0.45\linewidth}
        \centering
        \includegraphics[height=4cm]{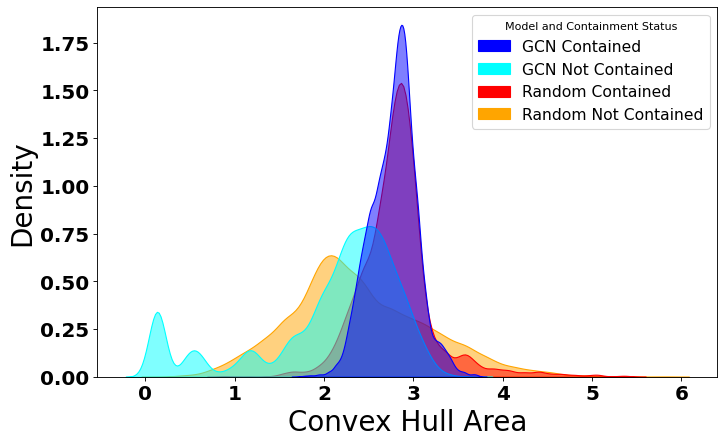}
        \caption{Conditional Density Plot showing containment metrics for 3 pursuers against 1 evader.}
    \end{subfigure} \hfill
    \begin{subfigure}[t]{0.45\linewidth}
        \centering
        \includegraphics[height=4cm]{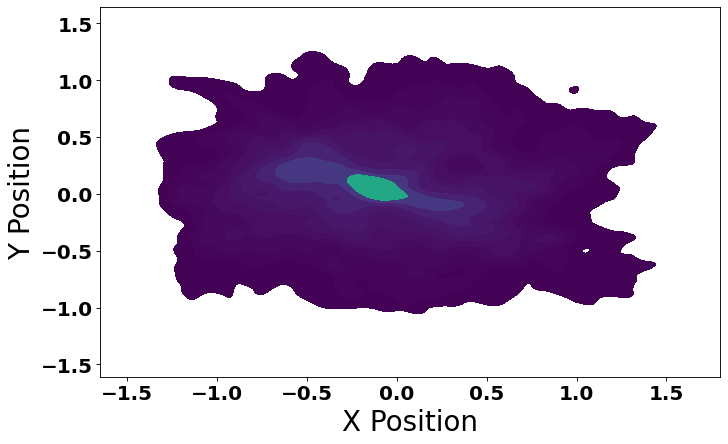}
        \caption{Heat map showing movement patterns with hot start strategies for 3 pursuers vs. 1 evader.}
    \end{subfigure} \hfill
    \begin{subfigure}[t]{0.45\linewidth}
        \centering
        \includegraphics[height=4cm]{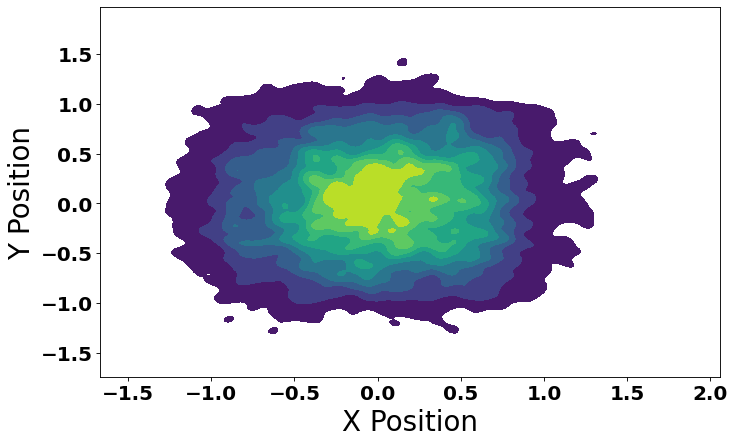}
        \caption{Heat map showing movement patterns with random start strategies for 3 pursuers vs. 1 evader.}
    \end{subfigure}
    \caption{Movement analysis for 3 Pursuers vs. 1 Evader games.}
    \label{fig:survival_analysis_3_pursuers}
\end{figure}

\subsubsection{1 Evader vs. 5 Pursuers Movement Analysis}
\begin{figure}[H]
    \centering
    \begin{subfigure}[t]{0.45\linewidth}
        \centering
        \includegraphics[height=4cm]{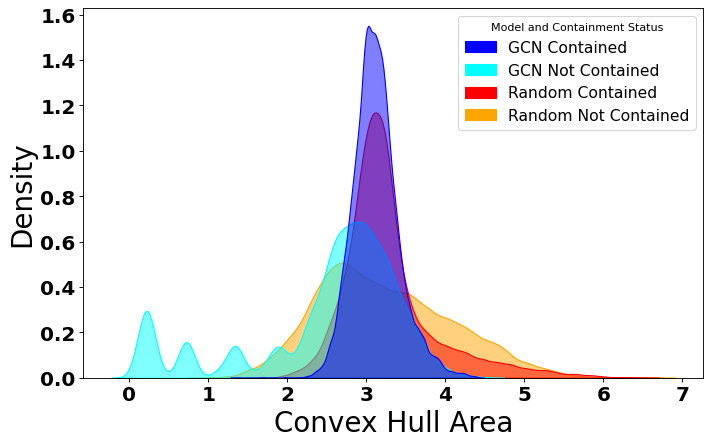}
        \caption{Conditional Density Plot showing containment metrics for 5 pursuers against 1 evader.}
    \end{subfigure} \hfill
    \begin{subfigure}[t]{0.45\linewidth}
        \centering
        \includegraphics[height=4cm]{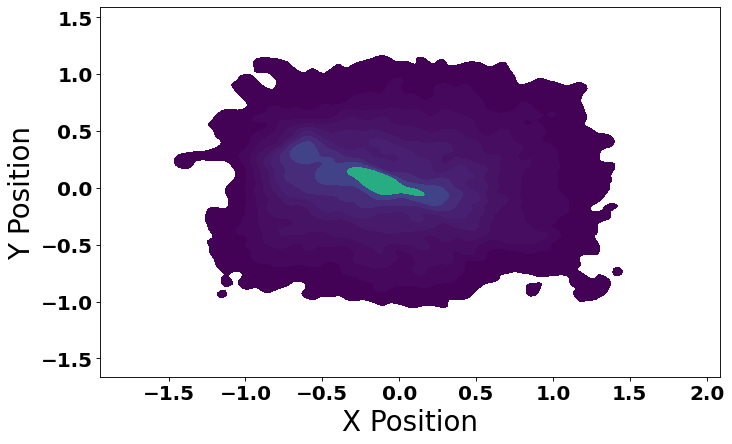}
        \caption{Heat map showing movement patterns with hot start strategies for 5 pursuers vs. 1 evader.}
    \end{subfigure} \hfill
    \begin{subfigure}[t]{0.45\linewidth}
        \centering
        \includegraphics[height=4cm]{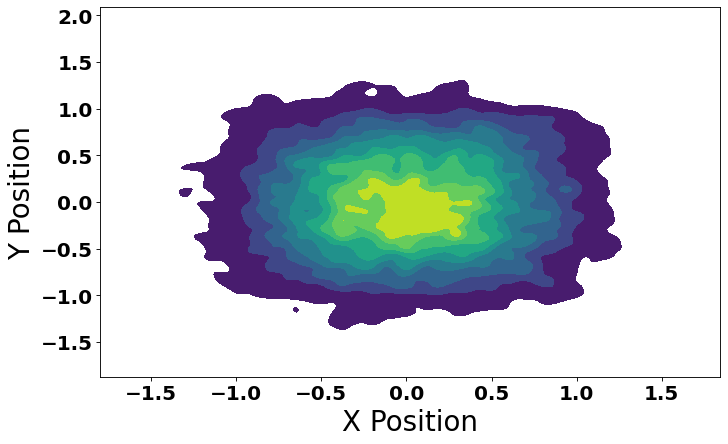}
        \caption{Heat map showing movement patterns with random start strategies for 5 pursuers vs. 1 evader.}
    \end{subfigure}
    \caption{Movement analysis for 5 Pursuers vs. 1 Evader games.}
    \label{fig:survival_analysis_5_pursuers}
\end{figure}
\subsection{Many vs. Many Pursuers Movement Analysis}
\subsubsection{4 Pursuers vs. 3 Evaders Movement Analysis}
\begin{figure}[H]
    \centering
    \begin{subfigure}[t]{0.45\linewidth}  
        \centering
        \includegraphics[height=4cm]{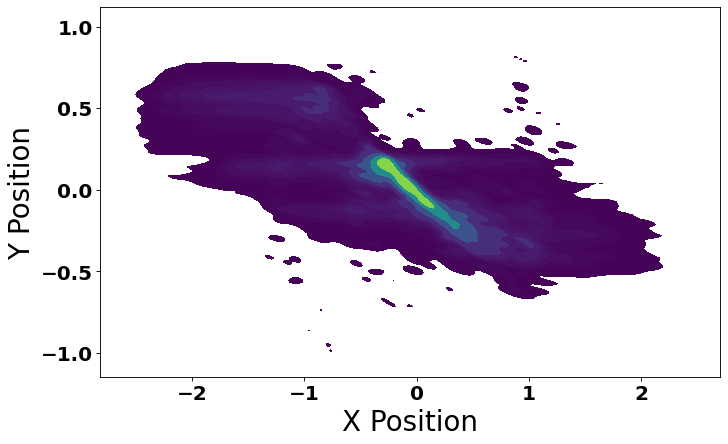}
        \caption{4 Pursuers vs. 3 Evaders GCN - Hot Starts heat map}
    \end{subfigure} \hfill
    \begin{subfigure}[t]{0.45\linewidth}
        \centering
        \includegraphics[height=4cm]{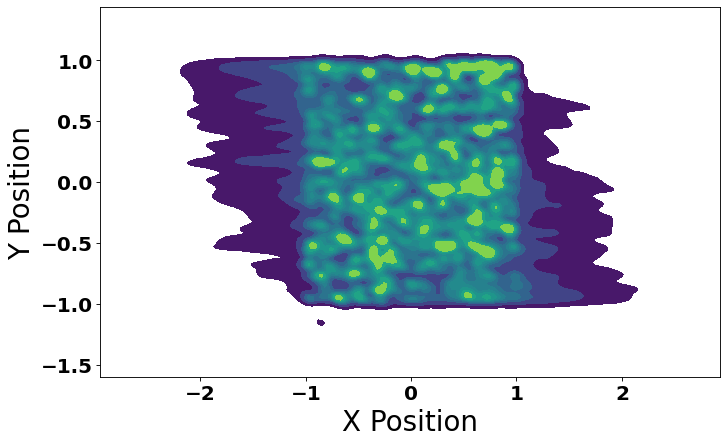}
        \caption{4 Pursuers vs. 3 Evaders Random Start heat map}
    \end{subfigure}
    \caption{Heat Maps for 4 Pursuers vs. 3 Evader Games. The brighter the color, the more activity in that area of movement.}
    \label{fig:your_figure_label_2}
\end{figure}
\subsubsection{4 Pursuers vs. 4 Evaders Movement Analysis}
\begin{figure}[H]
    \centering
    \begin{subfigure}[t]{0.45\linewidth}  
        \centering
        \includegraphics[height=4cm]{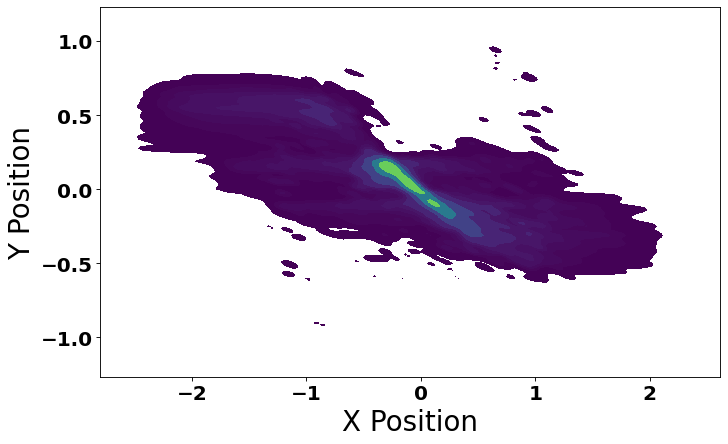}
        \caption{4 Pursuers vs. 4 Evaders GCN - Hot Starts heat map}
    \end{subfigure} \hfill
    \begin{subfigure}[t]{0.45\linewidth}
        \centering
        \includegraphics[height=4cm]{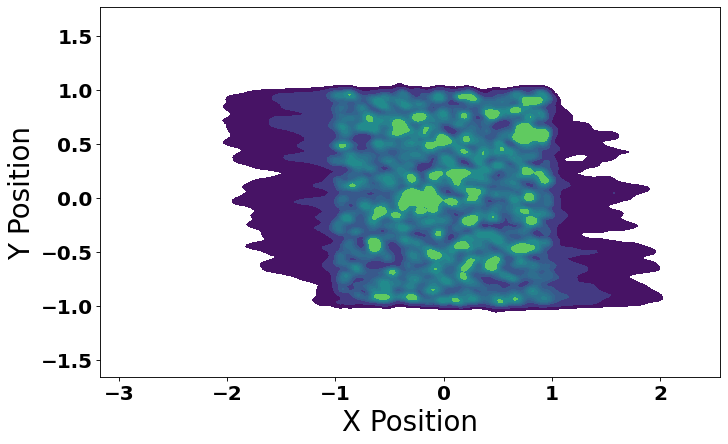}
        \caption{4 Pursuers vs. 4 Evaders Random Start heat map}
    \end{subfigure}
    \caption{Heat Maps for 4 Pursuers vs. 4 Evaders Games. The brighter the color, the more activity in that area of movement.}
    \label{fig:your_figure_label_2}
\end{figure}
\subsubsection{4 Pursuers vs. 5 Evaders Movement Analysis}
\begin{figure}[H]
    \centering
    \begin{subfigure}[t]{0.45\linewidth}  
        \centering
        \includegraphics[height=4cm]{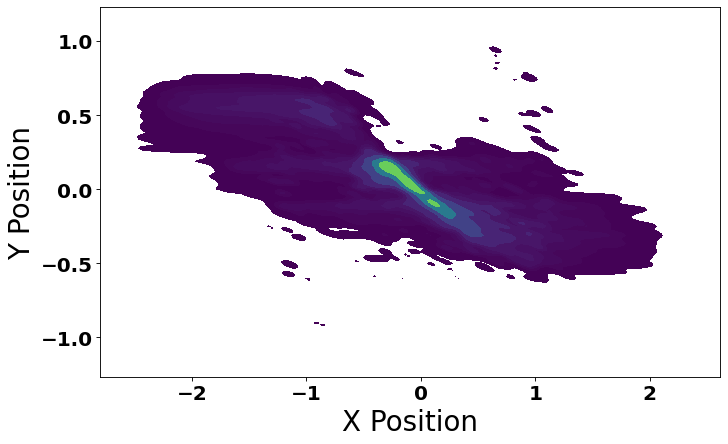}
        \caption{4 Pursuers vs. 5 Evaders GCN - Hot Starts heat map}
    \end{subfigure} \hfill
    \begin{subfigure}[t]{0.45\linewidth}
        \centering
        \includegraphics[height=4cm]{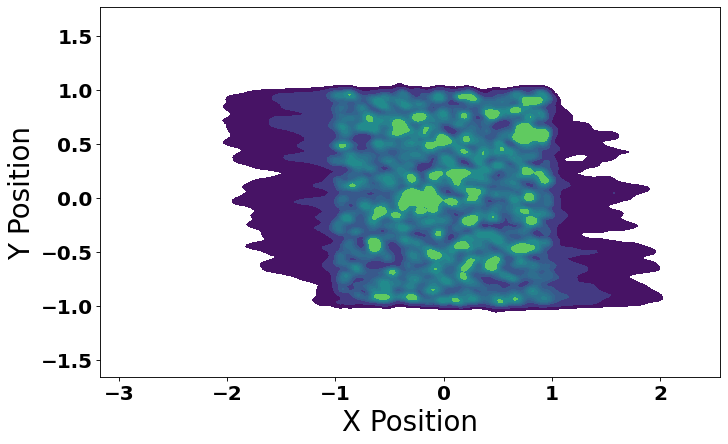}
        \caption{4 Pursuers vs. 5 Evaders Random Start heat map}
    \end{subfigure}
    \caption{Heat Maps for 4 Pursuers vs. 5 Evaders Games. The brighter the color, the more activity in that area of movement.}
    \label{fig:your_figure_label_2}
\end{figure}
\subsubsection{5 Pursuers vs. 2 Evaders Movement Analysis}
\begin{figure}[H]
    \centering
    \begin{subfigure}[t]{0.45\linewidth}  
        \centering
        \includegraphics[height=4cm]{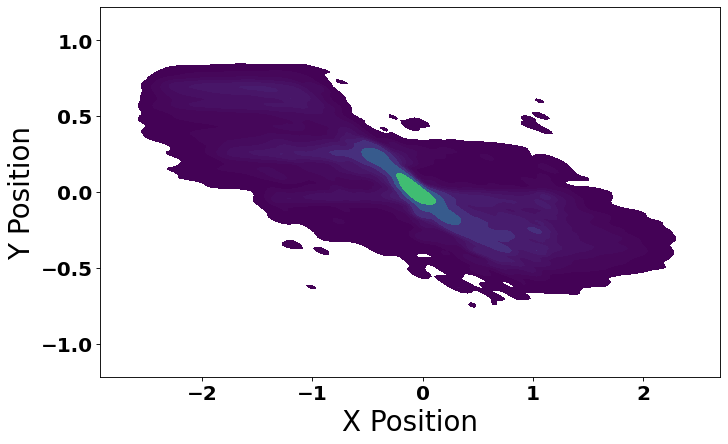}
        \caption{5 Pursuers vs. 2 Evaders GCN - Hot Starts heat map}
    \end{subfigure} \hfill
    \begin{subfigure}[t]{0.45\linewidth}
        \centering
        \includegraphics[height=4cm]{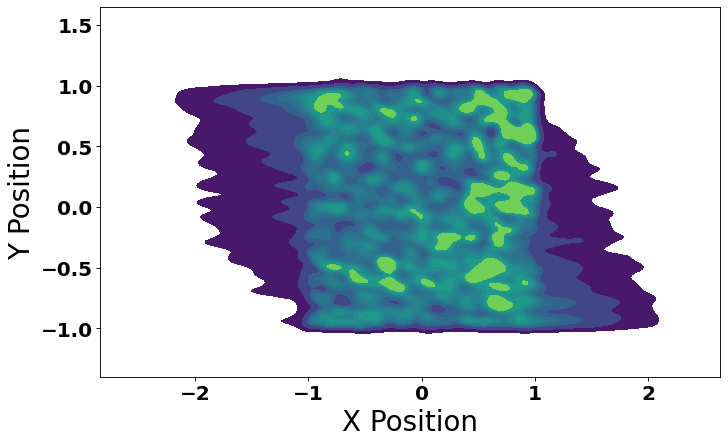}
        \caption{5 Pursuers vs. 2 Evaders Random Start heat map}
    \end{subfigure}
    \caption{Heat Maps for 5 Pursuers vs. 2 Evaders Games. The brighter the color, the more activity in that area of movement.}
    \label{fig:your_figure_label_2}
\end{figure}
\subsubsection{5 Pursuers vs. 3 Evaders Movement Analysis}
\begin{figure}[H]
    \centering
    \begin{subfigure}[t]{0.45\linewidth}  
        \centering
        \includegraphics[height=4cm]{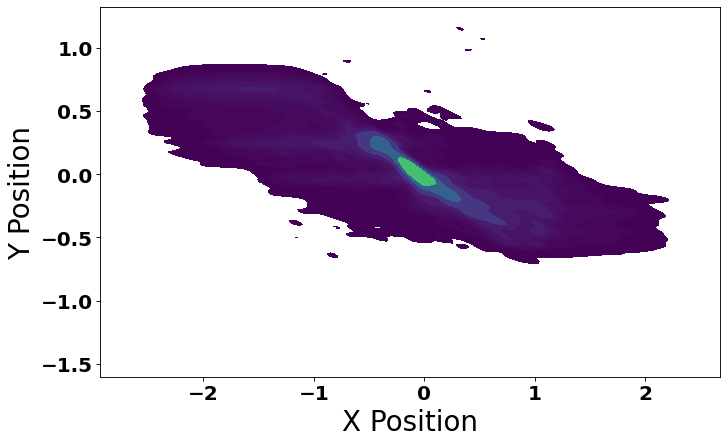}
        \caption{5 Pursuers vs. 3 Evaders GCN - Hot Starts heat map}
    \end{subfigure} \hfill
    \begin{subfigure}[t]{0.45\linewidth}
        \centering
        \includegraphics[height=4cm]{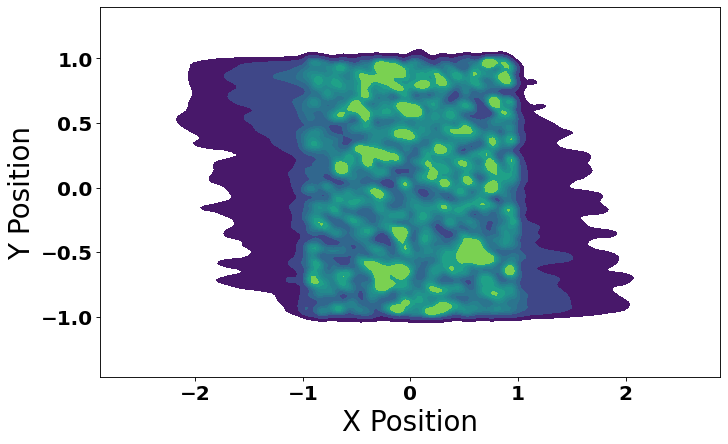}
        \caption{5 Pursuers vs. 3 Evaders Random Start heat map}
    \end{subfigure}
    \caption{Heat Maps for 5 Pursuers vs. 3 Evaders Games. The brighter the color, the more activity in that area of movement.}
    \label{fig:your_figure_label_2}
\end{figure}
\subsubsection{5 Pursuers vs. 4 Evaders Movement Analysis}
\begin{figure}[H]
    \centering
    \begin{subfigure}[t]{0.45\linewidth}  
        \centering
        \includegraphics[height=4cm]{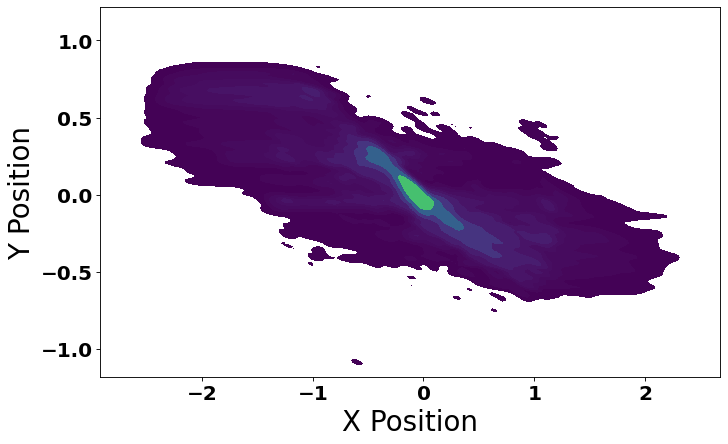}
        \caption{5 Pursuers vs. 4 Evaders GCN - Hot Starts heat map}
    \end{subfigure} \hfill
    \begin{subfigure}[t]{0.45\linewidth}
        \centering
        \includegraphics[height=4cm]{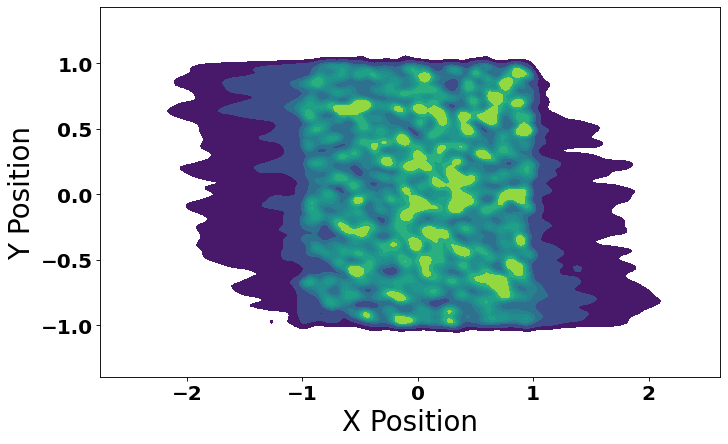}
        \caption{5 Pursuers vs. 4 Evaders Random Start heat map}
    \end{subfigure}
    \caption{Heat Maps for 5 Pursuers vs. 4 Evaders Games. The brighter the color, the more activity in that area of movement.}
    \label{fig:your_figure_label_2}
\end{figure}
\subsubsection{5 Pursuers vs. 5 Evaders Movement Analysis}
\begin{figure}[H]
    \centering
    \begin{subfigure}[t]{0.45\linewidth}  
        \centering
        \includegraphics[height=4cm]{AAMAS-2025-Formatting-Instructions-CCBY/5_v_4_HM_GCN.png}
        \caption{5 Pursuers vs. 5 Evaders GCN - Hot Starts heat map}
    \end{subfigure} \hfill
    \begin{subfigure}[t]{0.45\linewidth}
        \centering
        \includegraphics[height=4cm]{AAMAS-2025-Formatting-Instructions-CCBY/5_v_4_HM_rand.png}
        \caption{5 Pursuers vs. 5 Evaders Random Start heat map}
    \end{subfigure}
    \caption{Heat Maps for 5 Pursuers vs. 5 Evaders Games. The brighter the color, the more activity in that area of movement.}
    \label{fig:your_figure_label_2}
\end{figure}
\clearpage 
\bibliographystyle{plain} 
\bibliography{sample3}